\documentclass[12pt,letterpaper]{article}
\usepackage{amsmath,amssymb,array,calc,rotating,epsfig,psfrag,amscd, cite}

\makeatletter
\renewcommand\section{\@startsection {section}{1}{\z@}%
                                   {-3.5ex \@plus -1ex \@minus -.2ex}
                                   {2.3ex \@plus.2ex}%
                                   {\normalfont\large\bfseries}}
\renewcommand\subsection{\@startsection{subsection}{2}{\z@}%
                                     {-3.25ex\@plus -1ex \@minus -.2ex}%
                                     {1.5ex \@plus .2ex}%
                                     {\normalfont\bfseries}}
\makeatother

\let\non\nonumber

\let\S=\Sigma

\newcommand{\bea}{\begin{eqnarray}}
\newcommand{\eea}{\end{eqnarray}}
\newcommand{\be}{\begin{equation}}
\newcommand{\ee}{\end{equation}}

\newcommand{\m}{\mu}

\newcommand{\p}{\partial}

\newcommand{\C}[1]{$(\ref{#1})$}

\def\IZ{\relax\ifmmode\mathchoice
{\hbox{\cmss Z\kern-.4em Z}}{\hbox{\cmss Z\kern-.4em Z}}
{\lower.9pt\hbox{\cmsss Z\kern-.4em Z}} {\lower1.2pt\hbox{\cmsss
Z\kern-.4em Z}}\else{\cmss Z\kern-.4em Z}\fi}
\def\IR{\relax{\rm I\kern-.18em R}}

\def\one{{\hbox{ 1\kern-.8mm l}}}

\newlength{\bredde}
\def\slash#1{\settowidth{\bredde}{$#1$}\ifmmode\,\raisebox{.15ex}{/}
\hspace*{-\bredde} #1\else$\,\raisebox{.15ex}{/}\hspace*{-\bredde}
#1$\fi}

\newsavebox{\zzzbar}
\sbox{\zzzbar}
  {\setlength{\unitlength}{0.9em}
  \begin{picture}(0.6,0.7)
  \thinlines
  \put(0,0){\line(1,0){0.6}}
  \put(0,0.75){\line(1,0){0.575}}
  \multiput(0,0)(0.0125,0.025){30}{\rule{0.3pt}{0.3pt}}
  \multiput(0.2,0)(0.0125,0.025){30}{\rule{0.3pt}{0.3pt}}
  \put(0,0.75){\line(0,-1){0.15}}
  \put(0.015,0.75){\line(0,-1){0.1}}
  \put(0.03,0.75){\line(0,-1){0.075}}
  \put(0.045,0.75){\line(0,-1){0.05}}
  \put(0.05,0.75){\line(0,-1){0.025}}
  \put(0.6,0){\line(0,1){0.15}}
  \put(0.585,0){\line(0,1){0.1}}
  \put(0.57,0){\line(0,1){0.075}}
  \put(0.555,0){\line(0,1){0.05}}
  \put(0.55,0){\line(0,1){0.025}}
  \end{picture}}

\def\Im{{\rm Im ~}}
\def\Re{{\rm Re ~}}

\newcommand{\ena}{\end{eqnarray}}
\newcommand{\beqa}{\begin{eqnarray}}
\newcommand{\eeqa}{\end{eqnarray}}

\def\m{\mu}

\def\S{\Sigma}

\begin{document}
\begin{titlepage}

\begin{center}



\vskip 2 cm
{\Large \bf Eigenvalue equation for the modular graph $C_{a,b,c,d}$}\\
\vskip 1.25 cm { Anirban Basu\footnote{email address:
    anirbanbasu@hri.res.in} } \\
{\vskip 0.5cm  Harish--Chandra Research Institute, HBNI, Chhatnag Road, Jhusi,\\
Prayagraj 211019, India}

\end{center}

\vskip 2 cm

\begin{abstract}
\baselineskip=18pt

The modular graph $C_{a,b,c,d}$ on the torus is a three loop planar graph in which two of the vertices have coordination number four, while the others have coordination number two. We obtain an eigenvalue equation satisfied by $C_{a,b,c,d}$ for generic values of $a,b,c$ and $d$, where the source terms involve various modular graphs. This is obtained by varying the graph with respect to the Beltrami differential on the toroidal worldsheet. Use of several auxiliary graphs at various intermediate stages of the analysis is crucial in obtaining the equation. In fact, the eigenfunction is not simply $C_{a,b,c,d}$ but involves subtracting from it specific sums of squares of non--holomorphic Eisenstein series characterized by $a,b,c$ and $d$.

\end{abstract}

\end{titlepage}


\section{Introduction}

Understanding the properties of multiloop amplitudes in string theory is useful in obtaining terms in the effective action of the theory in a given background, as well as understanding non--perturbative duality symmetries at a quantitative level. In this paper, we shall be concerned with a specific $SL(2,\mathbb{Z})$ invariant integrand which arises in genus one string amplitudes which we denote as $C_{a,b,c,d}$. This is a modular graph function~\cite{DHoker:2015wxz} defined on the toroidal worldsheet for a fixed complex structure $\tau$. The links of this graph are given by scalar Green functions, while the vertices are given by the insertion points of vertex operators which are integrated over the toroidal worldsheet. $C_{a,b,c,d}$ is a three loop planar graph in which two of the vertices have coordination number four, while the others have coordination number two.     

To diagrammatically denote modular graphs, we denote the Green functions simply by lines. We often consider a chain of $a$ Green functions connected to each other such that all vertices apart from the ones at the end have coordination number two. This is denoted graphically by figure 1.   

\begin{figure}[ht]
\begin{center}
\[
\mbox{\begin{picture}(300,60)(0,0)
\includegraphics[scale=.6]{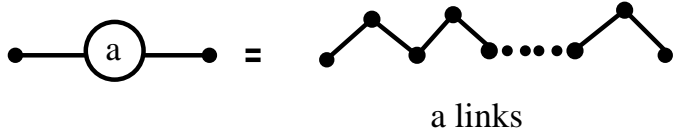}
\end{picture}}
\]
\caption{A chain with $a$ links}
\end{center}
\end{figure}

Thus $C_{a,b,c,d}$ is defined diagrammatically in figure 2. The detailed expression for this graph will be given later.  

\begin{figure}[ht]
\begin{center}
\[
\mbox{\begin{picture}(120,95)(0,0)
\includegraphics[scale=.65]{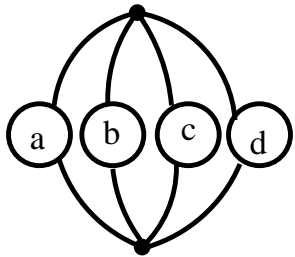}
\end{picture}}
\]
\caption{The modular graph $C_{a,b,c,d}$}
\end{center}
\end{figure}

Note that formally simply based on the algebraic formulae to be described later, $C_{a,b,c,d}$ is defined for arbitrary $a,b,c$ and $d$ provided the integrand has nice convergence properties, though for physically relevant cases that arise in string theory, $a,b,c$ and $d$ are positive integers. In fact, $C_{a,b,c,d}$ is a family of graphs characterized by $a,b,c$ and $d$. 

Such modular graphs arise in the low momentum expansion of string amplitudes which yield local terms in the effective action. The number of vertices and the number of links increase as one goes to higher point functions and higher orders in the derivative expansion respectively. These graphs have to be integrated over the truncated fundamental domain of $SL(2,\mathbb{Z})$, and the cutoff independent part of the final answer yields terms in the effective action~\cite{Green:1999pv,Green:2008uj}. In order to perform this analysis for various modular graphs, it is very useful to study their properties and obtain differential equations they satisfy on moduli space. Coupled with the asymptotic analysis of their behaviour near the cusp, one can obtain their contributions to the effective action. These issues have been discussed in various contexts in~\cite{DHoker:2015gmr,Basu:2015dqa,Basu:2015ayg,DHoker:2015wxz,Zerbini:2015rss,Basu:2016fpd,DHoker:2016mwo,Basu:2016xrt,Basu:2016kli,Basu:2016mmk,DHoker:2016quv,Kleinschmidt:2017ege,Brown:2017qwo,Basu:2017nhs,Basu:2017zvt,Broedel:2018izr,Zerbini:2018sox,Gerken:2018zcy,Gerken:2018jrq,DHoker:2019txf,Dorigoni:2019yoq,DHoker:2019xef,DHoker:2019mib,DHoker:2019blr}.

Thus obtaining differential equations which modular graphs satisfy is interesting and quite useful, and such equations have indeed been obtained for various graphs. Rather than considering specific graphs, it would be nice to obtain equations satisfied by families of such graphs, as they not only give insight into the structure of graphs with varying number of links in the same family, but this information is also useful in analyzing properties of such graphs at all orders in the $\alpha'$ expansion. Such eigenvalue equations have been obtained for the families of graphs $E_a$, $C_{a,b,c}$ and $M_{a,b,c,d,e,f}$~\cite{DHoker:2015gmr,Kleinschmidt:2017ege}, where these graphs are defined later. While the one loop graph $E_a$ satisfies Laplace equation on moduli space, the two loop graph $C_{a,b,c}$ and the three loop graph $M_{a,b,c,d,e,f}$ satisfy Poisson equations.   
The analysis is simple because these graphs do not have vertices with coordination number more than three. 

The analysis changes qualitatively and presents a richer structure once graphs with vertices with coordination number more than three are allowed. For this reason in this paper, we look at the family of graphs $C_{a,b,c,d}$ where two of the vertices have coordination number four. The structure of the source terms that arise in the Poisson equation we obtain is qualitatively different from that for $C_{a,b,c}$ and $M_{a,b,c,d,e,f}$.           

To obtain the differential equation satisfied by $C_{a,b,c,d}$ for generic $a,b,c$ and $d$, we perform the holomorphic and then the anti--holomorphic variations of the graph with respect to the Beltrami differential and manipulate the resulting expressions using elementary properties of the Green function to obtain the eigenvalue equation, along the lines of~\cite{Basu:2016kli}. This involves the use of auxiliary graphs~\cite{Basu:2016xrt} at various intermediate stages of the analysis. This yields the desired Poisson equation with source terms given by elementary modular graphs with links given by scalar Green functions.    

There are two major qualitative differences in the structure of the Poisson equation for $C_{a,b,c,d}$ compared to $C_{a,b,c}$ and $M_{a,b,c,d,e,f}$. First of all, for $C_{a,b,c}$ the source terms lie in the family $C_{a,b,c}$ with shifted labels, and similar is the case for $M_{a,b,c,d,ef}$ for generic values of the labels. However, for $C_{a,b,c,d}$ the source terms involve families of graphs other than $C_{a,b,c,d}$ as well. Secondly, in the Poisson equations for $C_{a,b,c}$ and $M_{a,b,c,d,ef}$, the eigenfunctions are $C_{a,b,c}$ and $M_{a,b,c,d,ef}$ respectively. However, the eigenfunction in the equation for $C_{a,b,c,d}$ is given by
\be \label{shift}C_{a,b,c,d} - E_{a+b} E_{c+d} - E_{a+c} E_{b+d} - E_{a+d} E_{b+c}.\ee  
Such shifts in the eigenfunction have been observed in various other cases as well, and the subtracted terms amount to removing contributions that are obtained by cutting open the parent graph in all possible ways. Thus what remains is really what should be thought of as the irreducible part of the parent graph\footnote{This has been termed ``primitive modular graph function'' in~\cite{DHoker:2016quv}. This phenomenon has also been observed at genus two~\cite{Basu:2018bde}, where the eigenfunction has a similar graphical interpretation. All this suggests this might be true in general.}. The terms that need to be subtracted from the parent graph arise very naturally in our graphical analysis. 
The strategy of using auxiliary graphs to obtain eigenvalue equations is general, hence we expect it to be useful to obtain such equations for other classes of modular graph functions, and more generally modular graph forms, as well.
     
We begin by first reviewing briefly various details of the scalar Green function as well as the variations of relevant quantities under the variations of Beltrami differentials on the worldsheet. After defining the various graphs, we then deduce the eigenvalue equation for $C_{a,b,c}$ using our method\footnote{This has already been deduced in~\cite{DHoker:2015gmr} using a different method. We revisit the eigenvalue equation with an eye towards generalizing it to $C_{a,b,c,d}$. The eigenvalue equation for $M_{a,b,c,d,e,f}$ has been obtained in~\cite{Kleinschmidt:2017ege} using the methods used in the present paper.}. Next we obtain the eigenvalue equation for $C_{a,b,c,d}$ for generic labels $a,b,c$ and $d$. The source terms in the Poisson equation can have vanishing or negative labels for certain choices of labels in the original graph. We discuss them in some detail and look at some simple examples.       

\section{The scalar Green function and varying the Beltrami differential}

We briefly review various properties of the conformally invariant scalar Green function that are relevant for our analysis.

The coordinate $z$ on the toroidal worldsheet $\S$ with complex structure $\tau$ lies in the region
\be -\frac{1}{2} \leq {\Re} z \leq \frac{1}{2}, \quad 0 \leq {\Im} z \leq \tau_2.\ee

The single valued scalar Green function $G(z,w)$ between points $z$ and $w$ on the toroidal worldsheet with complex structure $\tau$ is given by\cite{Lerche:1987qk,Green:1999pv}
\be \label{exp}G(z,w) = \frac{1}{\pi} \sum_{(m,n) \neq (0,0)} \frac{\tau_2}{\vert m\tau+ n\vert^2} e^{\pi[\bar{y}(m\tau+n) - y(m\bar\tau +n)]/\tau_2}\ee
where $y= z-w$.
Thus we have that
\be \int_\S d^2z G(z,w) =0\ee
and hence graphs with an endpoint vanish. Also we see from \C{exp} that one particle reducible graphs vanish. 

Using the single valuedness of the Green function, we can integrate by parts ignoring total derivatives. We also have that $\p_z G(z,w) = -\p_w G(z,w)$ using translational invariance. 

The Green function satisfies the eigenvalue equation
\be \label{Eigen}\p_z\overline\p_w G(z,w) = \pi \delta^2 (z-w) - \frac{\pi}{\tau_2}\ee
which we use very often in our analysis. 

Under the variation of the Beltrami differential $\m$, we can work out its action on the Green function and its variation using results in\cite{Verlinde:1986kw,DHoker:1988pdl}.
We get that~\cite{DHoker:2015gmr,Basu:2015ayg}
\be \label{hol}\p_\m G(z_1,z_2) = -\frac{1}{\pi} \int_\S d^2 w \p_w G(w,z_1) \p_w G(w,z_2),\ee
while $\overline\p_\m \p_\m G(z_1,z_2)=0$ upto contributions from the boundary of moduli space that are not relevant for our purposes\footnote{They become relevant for certain modular graph forms that arise, for example, in heterotic string theory~\cite{Ellis:1987dc,Abe:1988cq,Basu:2017nhs,Gerken:2018jrq}.}.

The $SL(2,\mathbb{Z})$ invariant Laplacian is given by
\be \Delta = \overline\p_\m \p_\m = 4\tau_2^2 \frac{\p^2}{\p \tau \p \overline\tau}.\ee

\begin{figure}[ht]
\begin{center}
\[
\mbox{\begin{picture}(250,80)(0,0)
\includegraphics[scale=.7]{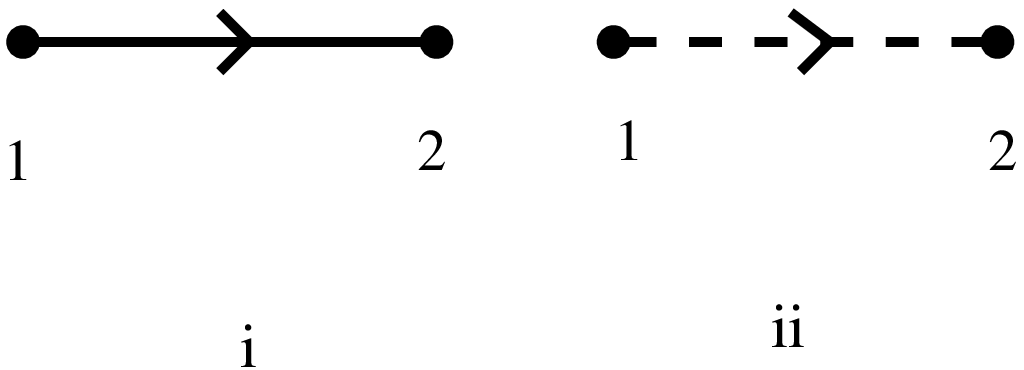}
\end{picture}}
\]
\caption{(i) $\p_{z_2} G(z_1,z_2) = -\p_{z_1} G(z_1,z_2)$,  (ii) $\overline\p_{z_2} G(z_1,z_2) = -\overline\p_{z_1} G(z_1,z_2)$}
\end{center}
\end{figure}

In the expressions for the various modular graphs in our analysis, all vertices are integrated with the measure $d^2 z/\tau_2$. In the intermediate steps, the number of $\tau_2$ factors in the integrand in the various graphs is determined in an obvious way from the formulae for the variations.       

\begin{figure}[ht]
\begin{center}
\[
\mbox{\begin{picture}(130,70)(0,0)
\includegraphics[scale=.6]{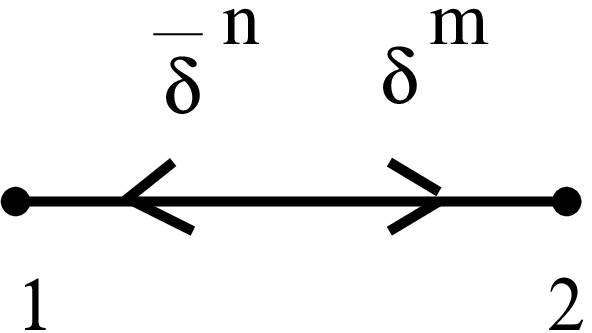}
\end{picture}}
\]
\caption{$\overline\p_{z_1}^n \p_{z_2}^m G(z_1,z_2)$}
\end{center}
\end{figure}

In the graphs, we denote Green functions simply by lines. For links having $\p_z G(z,w)$ or $\overline\p_z G(z,w)$, we denote them as shown in figure 3. We also need links having both holomorphic as well as anti--holomorphic derivatives of the Green function in our analysis, which is denoted by figure 4. The variation $\p_\m$ of the Green function is denoted as $\m$ next to the link, and similarly for its complex conjugate in the various graphs.    

\section{The various modular graph functions}

The simplest family of modular graphs involves $a$ links forming a closed chain given in figure 5. This one loop graph in which all vertices have coordination number two satisfies a simple differential equation on moduli space. 

\begin{figure}[ht]
\begin{center}
\[
\mbox{\begin{picture}(100,60)(0,0)
\includegraphics[scale=.55]{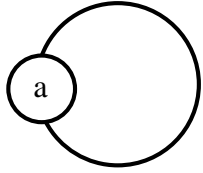}
\end{picture}}
\]
\caption{The modular graph $E_a$}
\end{center}
\end{figure}

To see this and also for later use, we define\footnote{Here $\S^a$ means $a$ copies of $\S$.}
\be {\mathcal{G}} (z_1,z_{a+1};a) \equiv \int_{\S^{a-1}}\prod_{i=2}^a \frac{d^2 z_i}{\tau_2} G(z_1,z_2) G(z_2,z_3) \ldots G(z_a,z_{a+1}).\ee
This is the chain with $a$ links depicted by figure 1, where the vertices at the end points have not been integrated over. 

Hence the modular graph $E_a$ given in figure 5 is given by
\be\label{Ea} E_a = \int_{\S} \frac{d^2 z}{\tau_2}{\mathcal{G}}(z,z;a) ,\ee 
which satisfies
\be \Delta E_a = a (a-1) E_a.\ee  

Another family of modular graphs which satisfies a simple differential equation is $C_{a,b,c}$ as given in figure 6. This two loop graph has two vertices with coordination number three, while the rest have coordination number two.  

\begin{figure}[ht]
\begin{center}
\[
\mbox{\begin{picture}(100,110)(0,0)
\includegraphics[scale=.6]{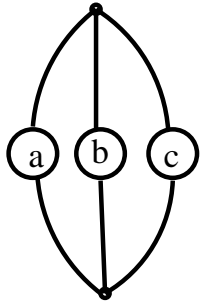}
\end{picture}}
\]
\caption{The modular graph $C_{a,b,c}$}
\end{center}
\end{figure}

We next consider the family of graphs $C_{a,b,c,d}$ given in figure 2 which is central to our analysis. For  $C_{a,b,c,d}$, the eigenvalue equation we shall obtain later is a second order differential equation with source terms. These involve graphs of the type $E_a$, $C_{a,b,c}$, $C_{a,b,c,d}$ as well as the three loop graphs $P_{a,b;c,d;e}$ and $M_{a,b,c,d,e,f}$ given in figure 7. While the graph $P_{a,b;c,d;e}$ has one vertex with coordination number four, the graph $M_{a,b,c,d,e,f}$ has no vertex with coordination number more then three. Thus we see that the eigenvalue equation for $C_{a,b,c,d}$ involves only elementary modular graphs with links given by Green functions.    

\begin{figure}[ht]
\begin{center}
\[
\mbox{\begin{picture}(320,150)(0,0)
\includegraphics[scale=.55]{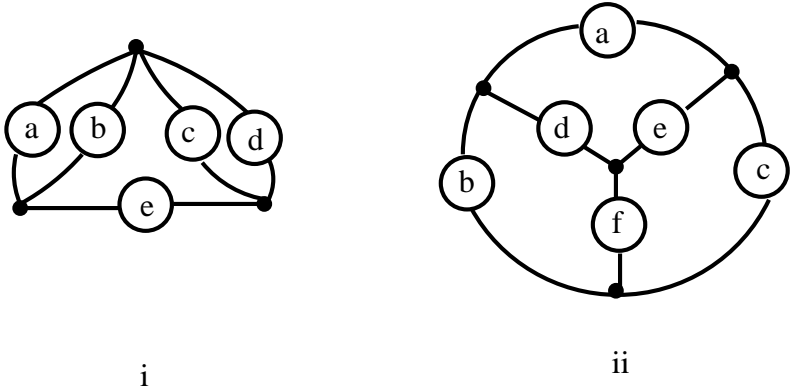}
\end{picture}}
\]
\caption{The modular graphs (i) $P_{a,b;c,d;e}$ and (ii) $M_{a,b,c,d,e,f}$}
\end{center}
\end{figure}

\section{Revisiting the eigenvalue equation for $C_{a,b,c}$}

We now consider the graph $C_{a,b,c}$ in figure 6, which is symmetric under interchange of $a,b$ and $c$. It is explicitly given by
\be \label{defCabc} C_{a,b,c} = \int_{\S^2} \prod_{i=1}^2 \frac{d^2 z_i}{\tau_2} \mathcal{G} (z_1,z_2;a) \mathcal{G} (z_1,z_2;b) \mathcal{G} (z_1,z_2;c).\ee
Thus such explicit expressions for the various graphs can be easily written down and we almost always simply express them diagrammatically for the sake of brevity.  

From \C{defCabc} performing the holomorphic variation of the Beltrami differential using \C{hol}, we get that 
\be \label{da}\p_\m C_{a,b,c} = a A(a;b,c) + b A(b;a,c) + c A(c;a,b),\ee
where $A(a;b,c) = A(a;c,b)$ is given in figure 8.

\begin{figure}[ht]
\begin{center}
\[
\mbox{\begin{picture}(100,90)(0,0)
\includegraphics[scale=.55]{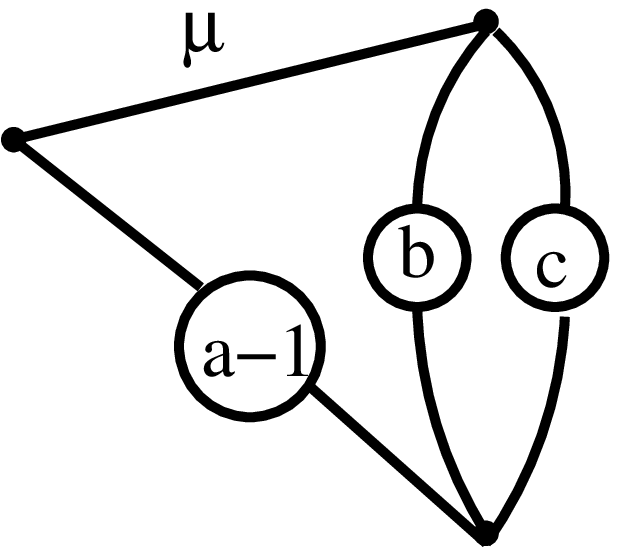}
\end{picture}}
\]
\caption{$A (a;b,c)$}
\end{center}
\end{figure} 

Thus further performing the anti--holomorphic variation of the Beltrami differential on \C{da}, we get that
\bea &&\Big[\overline\p_\m \p_\m -a(a-1) - b(b-1) - c(c-1)\Big]C_{a,b,c} \non \\ &&= ab \Big[B(a,b;c)+c.c.\Big]  + ac\Big[B(a,c;b) + c.c.\Big] + bc\Big[B(b,c;a)+c.c.\Big],\eea
where $B(a,b;c) = B(b,a;c)^*$ is given in figure 9. We have used \C{Eigen} at an intermediate step of our analysis to get the above expression. 

\begin{figure}[ht]
\begin{center}
\[
\mbox{\begin{picture}(120,100)(0,0)
\includegraphics[scale=.6]{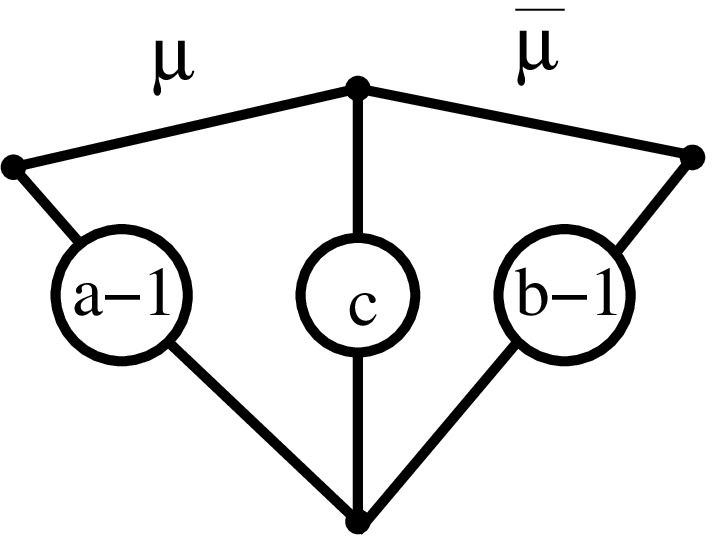}
\end{picture}}
\]
\caption{$B (a,b;c)$}
\end{center}
\end{figure}

We can calculate $B(a,b;c)$ independently to get that
\bea B(a,b;c) &=& - D(a,b;c-1) -\frac{1}{2} \Big[D(a,c-1;b) + D(c-1,b;a)\Big] \non \\ &&+\frac{1}{2} \Big[D(a-1,b;c)+ D(a,b-1;c)\Big],\eea 
where $D(a,b;c) = D(b,a;c)^*$ is given in figure 10. We also get that
\bea  D(a,b;c) + c.c. = C_{a,b+1,c} + C_{a+1,b,c} - C_{a+1,b+1,c-1}\eea
by a direct calculation. 

\begin{figure}[ht]
\begin{center}
\[
\mbox{\begin{picture}(120,85)(0,0)
\includegraphics[scale=.65]{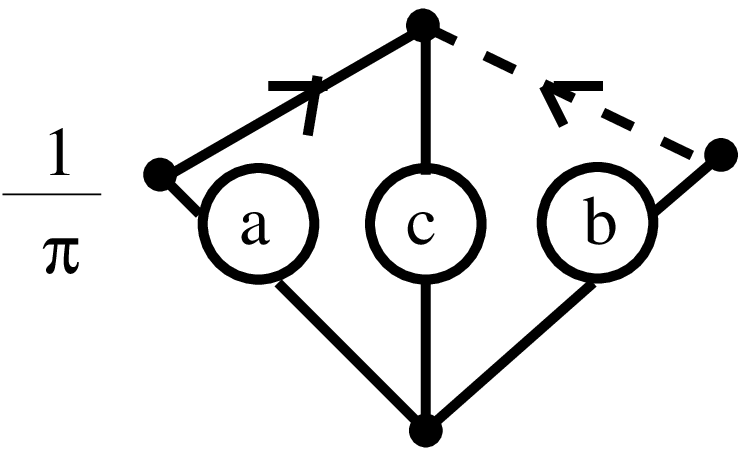}
\end{picture}}
\]
\caption{$D (a,b;c)$}
\end{center}
\end{figure}

Putting the various contributions together, we get that\footnote{In~\cite{DHoker:2015gmr}, this eigenvalue equation was deduced using the representation of the Green function as a lattice sum and manipulating the various expressions.}
\bea \label{eqn1}&&\Big[\Delta-a(a-1) - b(b-1) - c(c-1)\Big]C_{a,b,c} = f(a,b;c) + f(b,c;a) + f(a,c;b),\non \\ \eea
where
\bea f(a,b;c) &=& ab\Big[C_{a+1,b-1,c} + C_{a-1,b+1,c} + C_{a+1,b+1,c-2} -2 C_{a,b+1,c-1} -2 C_{a+1,b,c-1}\Big] \non \\ &=& f(b,a;c).\eea

In \C{eqn1}, though formally $C_{a,b,c}$ is defined for all $a,b$ and $c$ such that the integrand has nice convergence properties, we have that $a, b,c$ are positive integers for cases of interest in string amplitudes. Thus, on the right hand side, we can end up with the expressions $C_{a,b,0}$ and $C_{a,b,-1}$. These are analyzed in the appendix. 

\section{Deducing the structure of the eigenvalue equation for $C_{a,b,c,d}$}

We now consider the three loop modular graph $C_{a,b,c,d}$ in figure 2, which is symmetric under interchange of labels $a,b,c$ and $d$. Expressed as an integral over Green functions, it is given by
\be \label{defCabcd} C_{a,b,c,d} = \int_{\S^2} \prod_{i=1}^2 \frac{d^2 z_i}{\tau_2} \mathcal{G} (z_1,z_2;a) \mathcal{G} (z_1,z_2;b) \mathcal{G} (z_1,z_2;c) \mathcal{G} (z_1,z_2;d).\ee
To obtain the eigenvalue equation, we perform the action of the variation of the Beltrami differential on this graph. 

\subsection{The mixed variation of $C_{a,b,c,d}$}

\begin{figure}[ht]
\begin{center}
\[
\mbox{\begin{picture}(140,115)(0,0)
\includegraphics[scale=.55]{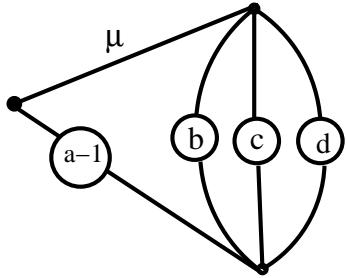}
\end{picture}}
\]
\caption{$A (a;b,c,d)$}
\end{center}
\end{figure}

Performing the holomorphic variation using \C{hol}, we get that
\be \label{A}\p_\m C_{a,b,c,d} = a A(a;b,c,d) + b A(b;a,c,d) + c A(c;a,b,d) + d A(d;a,b,c),\ee
where $A(a;b,c,d)$ is given in figure 11. Note that $A(a;b,c,d)$ is symmetric under interchange of $b,c$ and $d$. 

\begin{figure}[ht]
\begin{center}
\[
\mbox{\begin{picture}(140,90)(0,0)
\includegraphics[scale=.65]{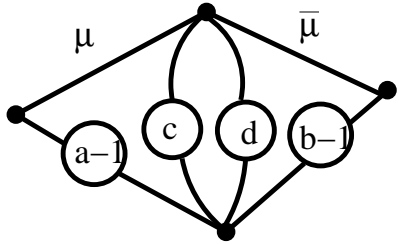}
\end{picture}}
\]
\caption{$B (a,b;c,d)$}
\end{center}
\end{figure}

We next perform the anti--holomorphic variation on the holomorphic variation of the graph given by \C{A} using the conjugate of \C{hol}. Using \C{Eigen}, this yields the equation for the mixed variation given by  
\bea \label{main}&&\Big[\overline\p_\m \p_\m -a(a-1) - b(b-1) - c(c-1) -d(d-1)\Big]C_{a,b,c,d}=ab \Big[B(a,b;c,d) + c.c.\Big]\non \\ &&+ ac \Big[B(a,c;b,d) + c.c.\Big]+ ad \Big[B(a,d;b,c) + c.c.\Big]+ bc \Big[B(b,c;a,d) + c.c.\Big]\non \\ &&+ bd \Big[B(b,d;a,c) + c.c.\Big]+ cd \Big[B(c,d;a,b) + c.c.\Big],\eea
where $B(a,b;c,d)$ is given in figure 12. Note that $B(a,b;c,d) = B(a,b;d,c) = B(b,a;c,d)^* = B(b,a;d,c)^*$.

Using \C{Eigen} and also integrating by parts in relevant terms, we get that
\bea \label{simplify} B(a,b;c,d) &=& G(a,c\vert d,b) +G(a,d\vert c,b) +\frac{1}{2} \Big[D(a,b-1;c,d)+ D(a-1,b;c,d)\Big] \non \\ &&-\frac{1}{2} \Big[D(c-1,b;a,d) + D(a,c-1;b,d) + D(d-1,b;a,c) + D(a,d-1;b,c)\Big]\non \\ && - D(a,b;c-1,d)- D(a,b;c,d-1),\eea
where $G(a,b\vert c,d) = G(d,c\vert b,a)^* = G(b-1,a+1 \vert c,d) = G(a,b \vert d+1,c-1) = G(b-1,a+1\vert d+1,c-1)$ is given in figure 13, and $D(a,b;c,d)= D(a,b;d,c) = D(b,a;c,d)^* = D(b,a;d,c)^*$ is given in figure 14. Thus $G(a,b\vert c,d)$ has two factors of $\p G$ and two factors of $\overline\p G$, while $D(a,b;c,d)$ has one factor of $\p G$ and one factor of $\overline\p  G$. 

\begin{figure}[ht]
\begin{center}
\[
\mbox{\begin{picture}(140,120)(0,0)
\includegraphics[scale=.65]{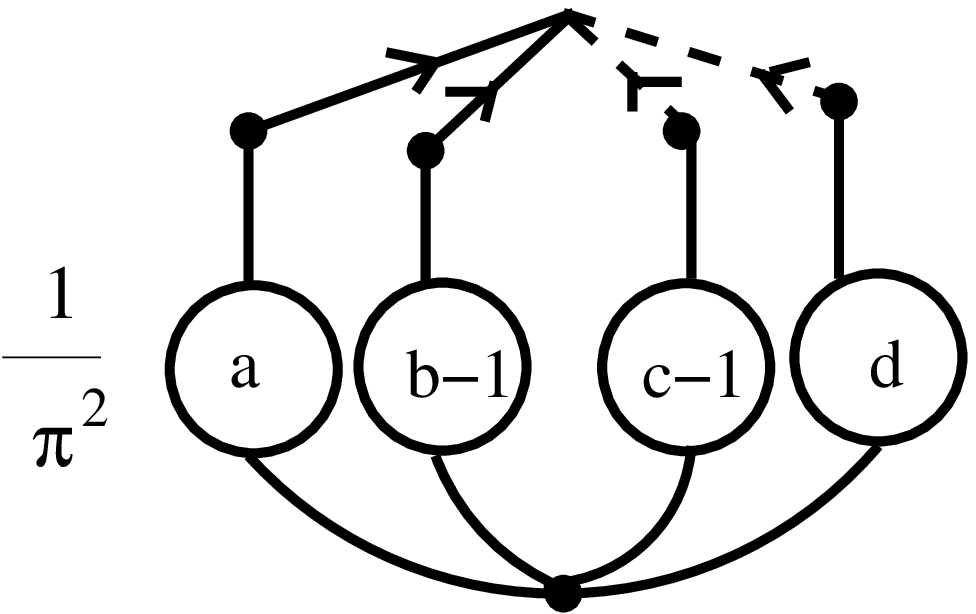}
\end{picture}}
\]
\caption{$G (a,b\vert c,d)$}
\end{center}
\end{figure}

\begin{figure}[ht]
\begin{center}
\[
\mbox{\begin{picture}(140,90)(0,0)
\includegraphics[scale=.75]{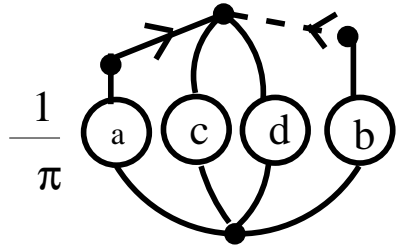}
\end{picture}}
\]
\caption{$D (a,b ; c,d)$}
\end{center}
\end{figure}

The terms of the type $D(a,b;c,d)$ in \C{simplify} can be further simplified to yield
\bea \label{defB} B(a,b;c,d) &=& G(a,c\vert d,b) +G(a,d\vert c,b)  + D(a,b-1;c,d)+ D(a-1,b;c,d) \non \\ &&- D(a,b;c-1,d) -D(a,b;c,d-1)-C_{a,b,c,d}\eea
which we now manipulate.

\begin{figure}[ht]
\begin{center}
\[
\mbox{\begin{picture}(190,120)(0,0)
\includegraphics[scale=.6]{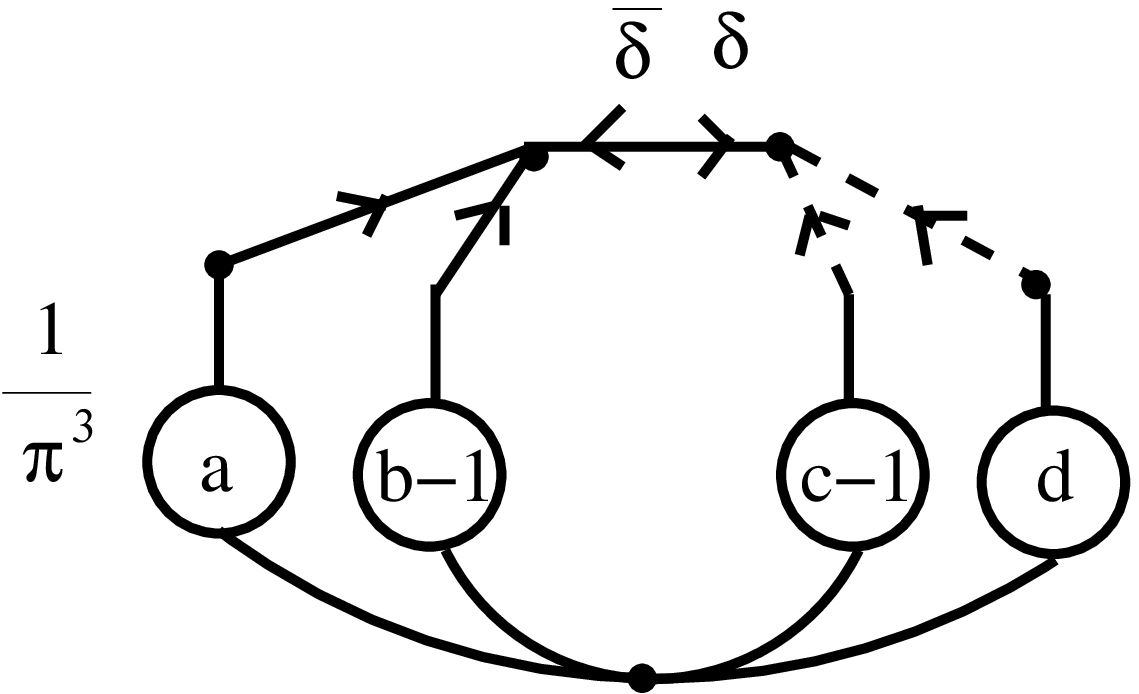}
\end{picture}}
\]
\caption{$H (a,b \vert c,d)$}
\end{center}
\end{figure}

\subsection{Simplifying $G(a,b\vert c,d)$}

First we consider terms of the type $G(a,b\vert c,d)$ in \C{defB}. To simplify $G(a,b \vert c,d)$, we consider the auxiliary graph $H(a,b\vert c,d)$ given in figure 15.
One of the links of this auxiliary graph is given by $\p_z \overline\p_w G (z,w)$. This is denoted by figure 4 for $m=n=1$. 

Simplifying it trivially using \C{Eigen} for the link of the form $\overline\p \p G$, we get that
\be \label{aux1}H(a,b\vert c,d) = G(a,b\vert c,d) - J(a+b-1) J(c+d-1)^*,\ee
where $J(a)$ is given in figure 16, and $\p_\m E_{a+1} = (a+1) J(a)$. 

\begin{figure}[ht]
\begin{center}
\[
\mbox{\begin{picture}(100,80)(0,0)
\includegraphics[scale=.6]{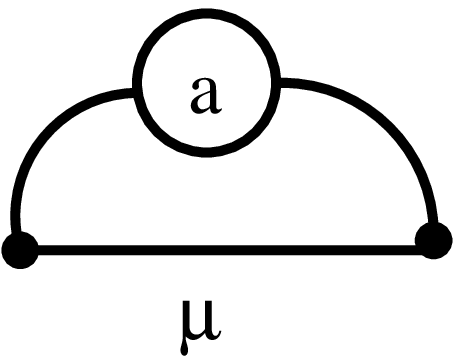}
\end{picture}}
\]
\caption{$J(a)$}
\end{center}
\end{figure}

Alternatively, we can evaluate $H(a,b\vert c,d)$ by first moving the $\overline\p$ in the link carrying both $\p$ and $\bar\p$ to the left and integrating by parts and using \C{Eigen}, and then moving the $\p$ to the right and again integrating by parts using \C{Eigen}. We get that
\be \label{aux2}H(a,b\vert c,d) = K(b-1,d;a,c-1)+K(b-1,c-1;a,d)+ K(a,d;b-1,c-1)+K(a,c-1;b-1,d), \ee
where $K(a,b;c,d) = K(b,a;d,c)^*$ is given in figure 17. Thus from \C{aux1} and \C{aux2} we get that
\bea \label{defG} G(a,b\vert c,d) &=& J(a+b-1)J(c+d-1)^* + K(b-1,d;a,c-1)\non \\ &&+K(b-1,c-1;a,d)+ K(a,d;b-1,c-1)+K(a,c-1;b-1,d). \non \\ \eea 

\begin{figure}[ht]
\begin{center}
\[
\mbox{\begin{picture}(210,80)(0,0)
\includegraphics[scale=.6]{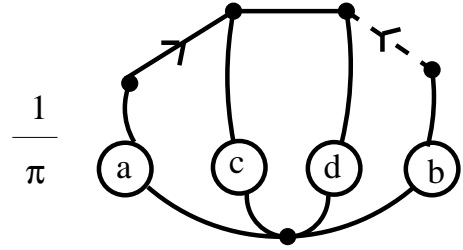}
\end{picture}}
\]
\caption{$K(a,b;c,d)$}
\end{center}
\end{figure}

This is a general strategy we shall use at various stages of our calculation. Often we obtain expressions in which there are factors of $\p G$ and $\overline\p G$ and we would like to simplify them to either get rid of the derivatives or put them in a form of diagrams which have derivatives in a useful way. However simply integrating by parts does not help in achieving our goal. We then introduce an auxiliary graph~\cite{Basu:2016xrt} which reduces to the graph under consideration trivially using \C{Eigen}. On the other hand, the auxiliary graph has to be judiciously chosen such that it can be independently simplified to produce graphs of the desired form. From now onwards, we shall simply mention the auxiliary graphs and write down the equality which is obtained following the steps mentioned above.    

Let us consider the contribution to the right hand side of \C{main} involving only the $J(p) J(q)^*$ terms that arise using \C{defG} in \C{defB}. This is equal to $X_{a,b,c,d}$ which is defined by
\bea X_{a,b,c,d} &=& (a+b)(c+d) \Big[J(a+b-1) J(c+d-1)^* + c.c.\Big] \non \\ &&+ (a+c)(b+d)\Big[J(a+c-1)J(b+d-1)^*+ c.c.\Big] \non \\ &&+ (a+d)(b+c)\Big[J(a+d-1)J(b+c-1)^* + c.c.\Big].\eea  

These contributions along with $\Delta C_{a,b,c,d}$ can be very usefully combined together by defining the combination
\be Y_{a,b,c,d} = C_{a,b,c,d} - E_{a+b} E_{c+d} - E_{a+c} E_{b+d} - E_{a+d} E_{b+c},\ee
using the identity
\bea \Delta Y_{a,b,c,d} &=& \Delta C_{a,b,c,d} - X_{a,b,c,d} - \Big[(a+b)(a+b-1)+ (c+d)(c+d-1)\Big] E_{a+b} E_{c+d} \non \\ &&- \Big[(a+c)(a+c-1)+ (b+d)(b+d-1)\Big] E_{a+c} E_{b+d} \non \\ && - \Big[(a+d)(a+d-1)+ (b+c)(b+c-1)\Big] E_{a+d} E_{b+c}.\eea

\begin{figure}[ht]
\begin{center}
\[
\mbox{\begin{picture}(120,100)(0,0)
\includegraphics[scale=.6]{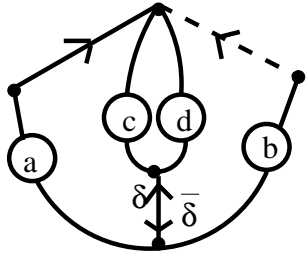}
\end{picture}}
\]
\caption{$P_1 (a,b;c,d)$}
\end{center}
\end{figure}

Thus \C{main} gives us that
\bea \label{eqn2}&&\Delta Y_{a,b,c,d} + \Big[(a+b)(a+b-1)+ (c+d)(c+d-1)\Big] E_{a+b} E_{c+d} \non \\&& + \Big[(a+c)(a+c-1)+ (b+d)(b+d-1)\Big] E_{a+c} E_{b+d} \non \\  && + \Big[(a+d)(a+d-1)+ (b+c)(b+c-1)\Big] E_{a+d} E_{b+c} \non \\ && -\Big[a(a-1)+b(b-1)+c(c-1)+d(d-1) -2(ab+ac+ad+bc+bd+cd)\Big]C_{a,b,c,d}\non \\
 &&= f(a,b;c,d) + f(a,c;b,d) +f(a,d;b,c) + f(b,c;a,d) + f(b,d;a,c)+ f(c,d;a,b),\non \\ \eea
where $f(a,b;c,d)$ only involves terms of the form $K(a,b;c,d)$ and $D(a,b;c,d)$, and is given by 
\bea \label{deff}f(a,b;c,d) &=& ab\Big[ K(c-1,b;a,d-1)+ K(d-1,b;a,c-1)+K(c-1,d-1;a,b) \non \\ &&+ K(d-1,c-1;a,b)+K(a,b;c-1,d-1)+ K(a,b;d-1,c-1) \non \\ && + K(a,c-1;d-1,b)+ K(a,d-1;c-1,b)  + D(a-1,b;c,d) \non \\ &&+ D(a,b-1;c,d) - D(a,b;c-1,d) - D(a,b;c,d-1)+ c.c.\Big] \non \\ &=& f(b,a;c,d) = f(a,b;d,c) = f(b,a;d,c).\eea

We now show that both terms of the form $D(a,b;c,d) + c.c.$ as well as $K(a,b;c,d)+ c.c.$ can be manipulated to yield graphs that are independent of derivatives. Then \C{eqn2} leads to the desired eigenvalue equation involving $C_{a,b,c,d}$ for generic values of $a,b,c$ and $d$.    

\subsection{Simplifying $D(a,b,c,d)$}

First let us consider $D(a,b;c,d)$. We start with the auxiliary graph $P_1 (a,b;c,d)$ given in figure 18. Evaluating it trivially, and alternatively by moving $\p$ and $\bar\p$ along the links forming the closed loop without derivatives and using \C{Eigen}, we get that   
\bea D(a,b;c,d) &=& E_{a+b+1}E_{c+d} + Q_1 (a,b;c-1,d) + Q_1 (a,b;d-1,c) \non \\ &&- Q_2 (a,c\vert d,b)- Q_2 (a,d\vert c,b),\eea
where $Q_1 (a,b;c,d)$ and $Q_2 (a,b \vert c,d)$ are given in figures 19 and 20 respectively. Note that $Q_1 (a,b;c,d) = Q_1 (a,b;d,c) = Q_1 (b,a;c,d)^* = Q_1 (b,a;d,c)^*$and $Q_2 (a,b\vert c,d) = Q_2 (d,c\vert b,a)^* =  Q_2 (b-1,a+1 \vert d+1,c-1)$.

\begin{figure}[ht]
\begin{center}
\[
\mbox{\begin{picture}(150,90)(0,0)
\includegraphics[scale=.6]{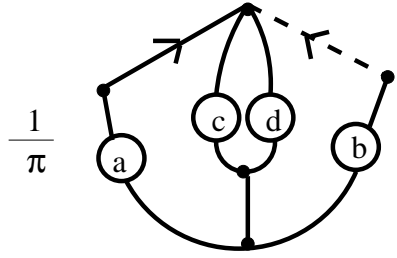}
\end{picture}}
\]
\caption{$Q_1 (a,b;c,d)$}
\end{center}
\end{figure}

\begin{figure}[ht]
\begin{center}
\[
\mbox{\begin{picture}(210,110)(0,0)
\includegraphics[scale=.55]{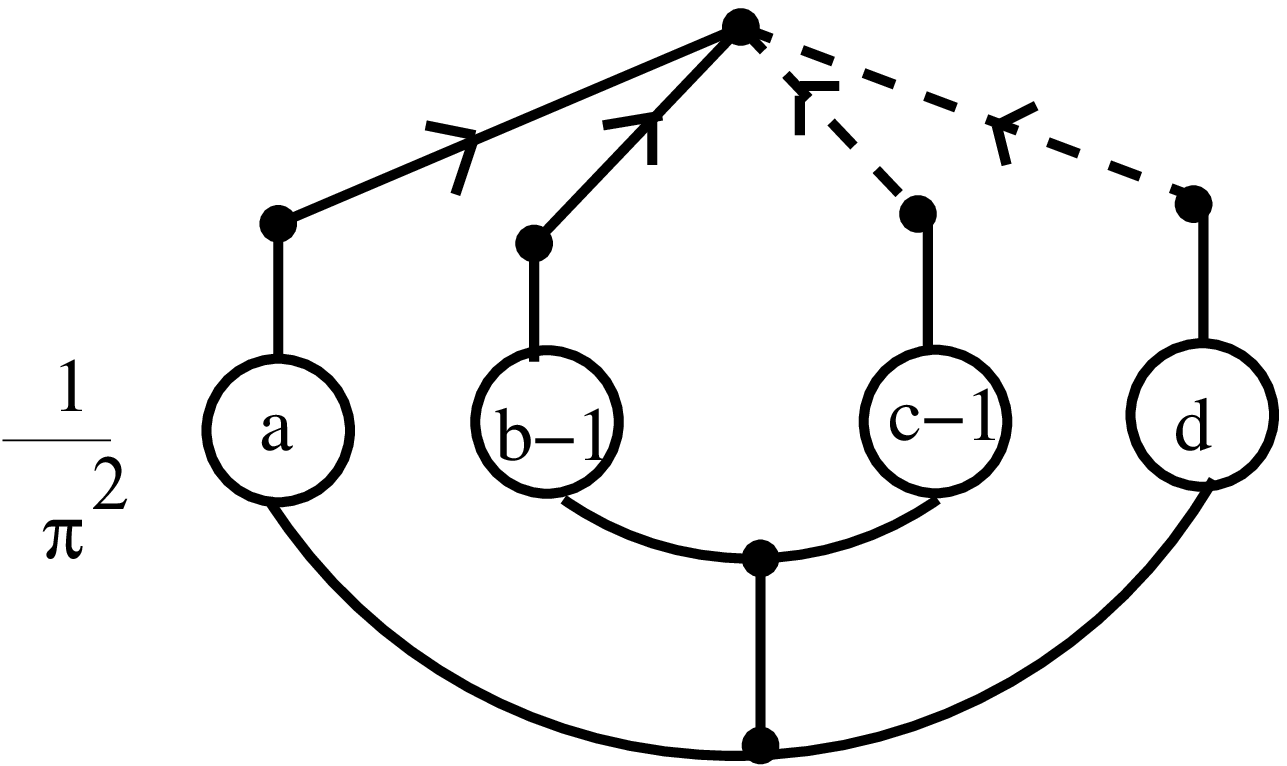}
\end{picture}}
\]
\caption{$Q_2 (a,b\vert c,d)$}
\end{center}
\end{figure}

\begin{figure}[ht]
\begin{center}
\[
\mbox{\begin{picture}(160,110)(0,0)
\includegraphics[scale=.5]{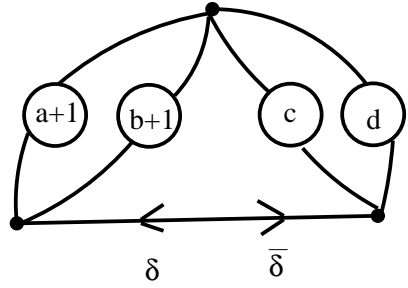}
\end{picture}}
\]
\caption{$P_2 (a+1,b+1;c,d)$}
\end{center}
\end{figure}

\begin{figure}[ht]
\begin{center}
\[
\mbox{\begin{picture}(170,130)(0,0)
\includegraphics[scale=.5]{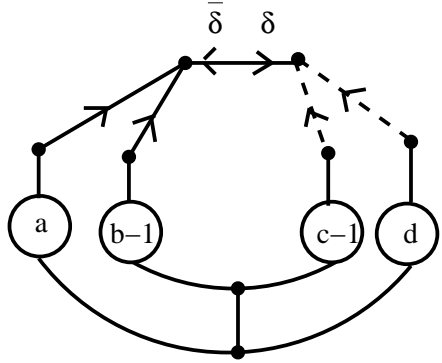}
\end{picture}}
\]
\caption{$P_3 (a,b;c,d)$}
\end{center}
\end{figure}

We now consider the graph $Q_1 (a,b;c,d)$. Starting with the auxiliary graph $P_2 (a+1,b+1;c,d)$ given in figure 21, we get that
\bea \label{q1}Q_1(a,b;c,d) + c.c. = E_{c+d} E_{a+b+2} - C_{a+1,b+1,c,d} + P_{a,b+1;c,d;1} + P_{a+1,b;c,d;1},\eea
and thus the graphs on the right hand side have no derivatives. 

Note that $P_{a,b;c,d;e}$ depicted by figure 7 is given by
\be \label{defP}P_{a,b;c,d;e} = \int_{\S^3} \prod_{i=1}^3 \frac{d^2 z_i}{\tau_2} \mathcal{G}(z_1,z_2;a) \mathcal{G}(z_1,z_2;b) \mathcal{G}(z_2,z_3;c) \mathcal{G} (z_2,z_3;d) \mathcal{G} (z_1,z_3;e)\ee
when expressed as an integral over factors of scalar Green functions.  

\begin{figure}[ht]
\begin{center}
\[
\mbox{\begin{picture}(180,100)(0,0)
\includegraphics[scale=.65]{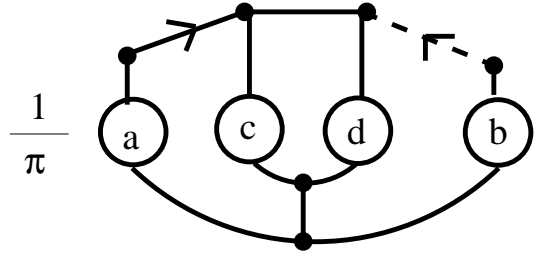}
\end{picture}}
\]
\caption{$Q_3 (a,b;c,d)$}
\end{center}
\end{figure}

\begin{figure}[ht]
\begin{center}
\[
\mbox{\begin{picture}(200,90)(0,0)
\includegraphics[scale=.55]{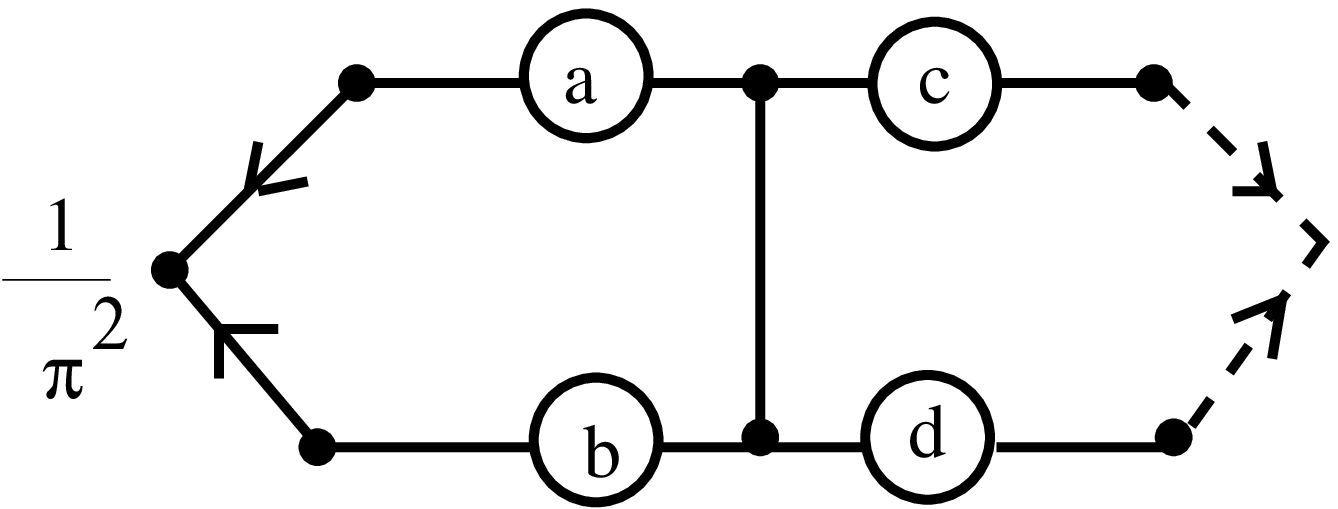}
\end{picture}}
\]
\caption{$Q_4 (a,b\vert c,d)$}
\end{center}
\end{figure}

We next consider the graph $Q_2 (a,b \vert c,d)$. We now start  with the auxiliary graph $P_3 (a,b\vert c,d)$ given in figure 22. Again to simplify it we proceed as before by evaluating it trivially, and then evaluating it again by moving around the derivatives and using \C{Eigen}. Equating the two expressions thus obtained gives us the required relation\footnote{This strategy of simplifying various graphs involving derivatives of Green functions using auxiliary graphs has been used very effectively, for example, in~\cite{Basu:2016mmk}. This involves joining and splitting of links in various graphs and seems to be a very useful technique in general.}. 

Thus we get that
\bea \label{Q2}Q_2 (a, b\vert c,d) &=& Q_3 (a,d; b-1,c-1) + Q_3 (b-1,c-1;a,d) \non \\ &&-\frac{1}{2}\Big[ Q_3 (a-1,d;b,c-1) + Q_3 (b-2,c-1;a+1,d) + Q_3 (a,d-1;b-1,c) \non \\ &&+ Q_3 (b-1,c-2;a,d+1)\Big] + Q_4 (a,b-1\vert d,c-1) \non \\ &&+\frac{1}{2} \Big[ Q_5 (a+1,b-1;c-1,d) + Q_5 (b,a ; d,c-1) + Q_5 (c,d;a,b-1)^* \non \\ &&+ Q_5 (d+1,c-1;b-1,a)^*\Big],\eea
where $Q_3 (a,b;c,d) = Q_3 (b,a;d,c)^*$, $Q_4 (a,b \vert c,d)$ and $Q_5 (a,b;c,d)$ are defined in figures 23, 24 and 25 respectively. 

\begin{figure}[ht]
\begin{center}
\[
\mbox{\begin{picture}(200,90)(0,0)
\includegraphics[scale=.55]{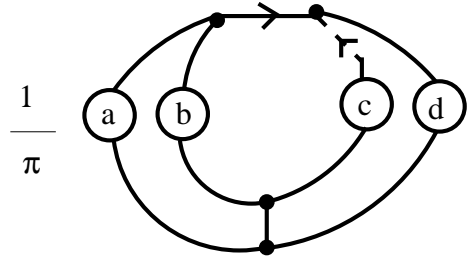}
\end{picture}}
\]
\caption{$Q_5 (a,b; c,d)$}
\end{center}
\end{figure}

\begin{figure}[ht]
\begin{center}
\[
\mbox{\begin{picture}(150,90)(0,0)
\includegraphics[scale=.7]{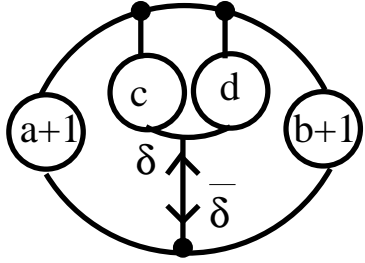}
\end{picture}}
\]
\caption{$P_4 (a,b; c,d)$}
\end{center}
\end{figure}

In order to calculate $Q_3 (a,b;c,d) + c.c.$, we consider the auxiliary graph $P_4 (a,b;c,d)$ given in figure 26, which gives us 
\bea \label{Q3}Q_3 (a,b;c,d) + c.c. = C_{1,a+b+2,c+d} - P_{a+1,c;b+1,d;1} + M_{1,a,b+1,c,d,1} + M_{1,a+1,b,c,d,1}.\eea 

Similarly, we get that
\bea \label{Q4}Q_4 (a,b\vert c,d) + c.c. &=& 4 E_{a+b+c+d+3} + C_{1,a+b,c+d+2} + C_{1,a+b+2,c+d}\non \\ &&- E_{a+b+2} E_{c+d+1} - E_{a+b+1} E_{c+d+2}.\eea

We also get that
\be \label{Q5}Q_5 (a,b;c,d) + c.c. = P_{a,d;b,c+1;1} - C_{1,a+b,c+d+1} + M_{1,c,b,d,a,1} - M_{1,c+1,b,d-1,a,1}.\ee
 
Note that $M_{a,b,c,d,e,f}$ depicted by figure 7 is given by
\be M_{a,b,c,d,e,f} = \int_{\S^4} \prod_{i=1}^4 \frac{d^2 z_i}{\tau_2} \mathcal{G} (z_1,z_2;a)\mathcal{G} (z_1,z_3;d) \mathcal{G} (z_1,z_4;b) \mathcal{G} (z_2,z_3;e) \mathcal{G} (z_2,z_4;c) \mathcal{G} (z_3,z_4;f)\ee
when expressed as an integral over factors of scalar Green functions.
At various places, we have used the relations\footnote{See~\cite{Basu:2014hsa} for example, for a relevant discussion.}
\be M_{a,b,c,d,e,f} = M_{a,e,d,c,b,f} = M_{a,d,e,b,c,f} = M_{a,c,b,e,d,f}\ee
to simplify the various expressions.

Thus, using \C{Q3}, \C{Q4} and \C{Q5} in \C{Q2}, we see that $Q_2 (a, b\vert c,d) + c.c.$ yields graphs with no derivatives. Along with \C{q1}, we conclude that $D(a,b;c,d) + c.c.$ yields graphs with no derivatives, leading to a simplification in the structure of the source terms. 

\begin{figure}[ht]
\begin{center}
\[
\mbox{\begin{picture}(160,110)(0,0)
\includegraphics[scale=.6]{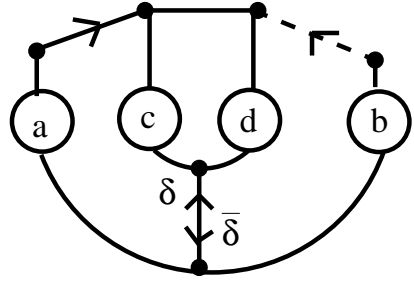}
\end{picture}}
\]
\caption{$R_1 (a,b; c,d)$}
\end{center}
\end{figure}

\begin{figure}[ht]
\begin{center}
\[
\mbox{\begin{picture}(180,110)(0,0)
\includegraphics[scale=.55]{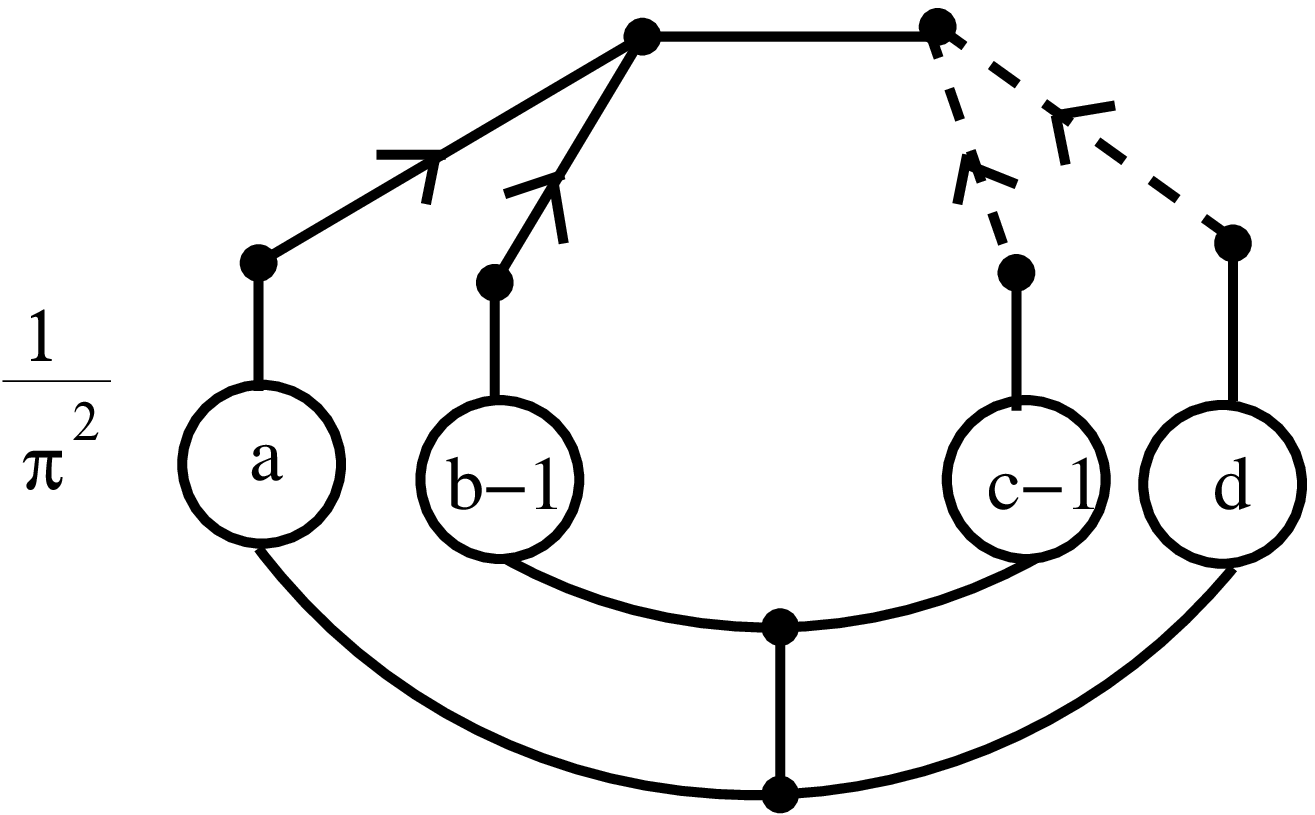}
\end{picture}}
\]
\caption{$S_1 (a,b\vert c,d)$}
\end{center}
\end{figure}

\begin{figure}[ht]
\begin{center}
\[
\mbox{\begin{picture}(180,80)(0,0)
\includegraphics[scale=.75]{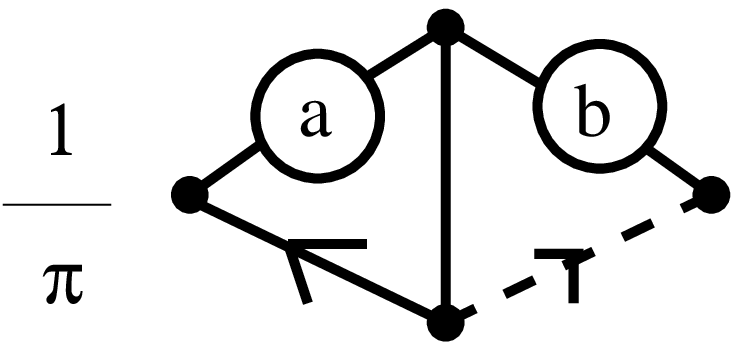}
\end{picture}}
\]
\caption{$S_2 (a;b)$}
\end{center}
\end{figure}

\subsection{Simplifying $K(a,b,c,d)$}

We next consider $K(a,b;c,d)$. Starting with the auxiliary graph $R_1 (a,b;c,d)$ given in figure 27 and proceeding as before, we get that
\bea &&K(a,b;c,d) = C_{1,a+b+1,c+d} + Q_3 (a,b;c-1,d) + Q_3 (a,b;c,d-1) - S_1 (a,c\vert d,b) \non \\ &&+ S_2 (a+c;b+d) - M_{1,a,b,c,d,1} - Q_1 (a,b;c,d) + Q_5 (c,a;b,d) + Q_5 (d,b;a,c)^*,\eea
where $S_1 (a, b\vert c,d)$ and $S_2 (a;b)$ are given in figures 28 and 29 respectively. Note that $S_1 (a,b\vert c,d) = S_1 (d,c\vert b,a)^* = S_1 (b-1,a+1\vert d+1,c-1)$, while $S_2 (a;b) = S_2 (b;a)^*$.

\begin{figure}[ht]
\begin{center}
\[
\mbox{\begin{picture}(180,150)(0,0)
\includegraphics[scale=.55]{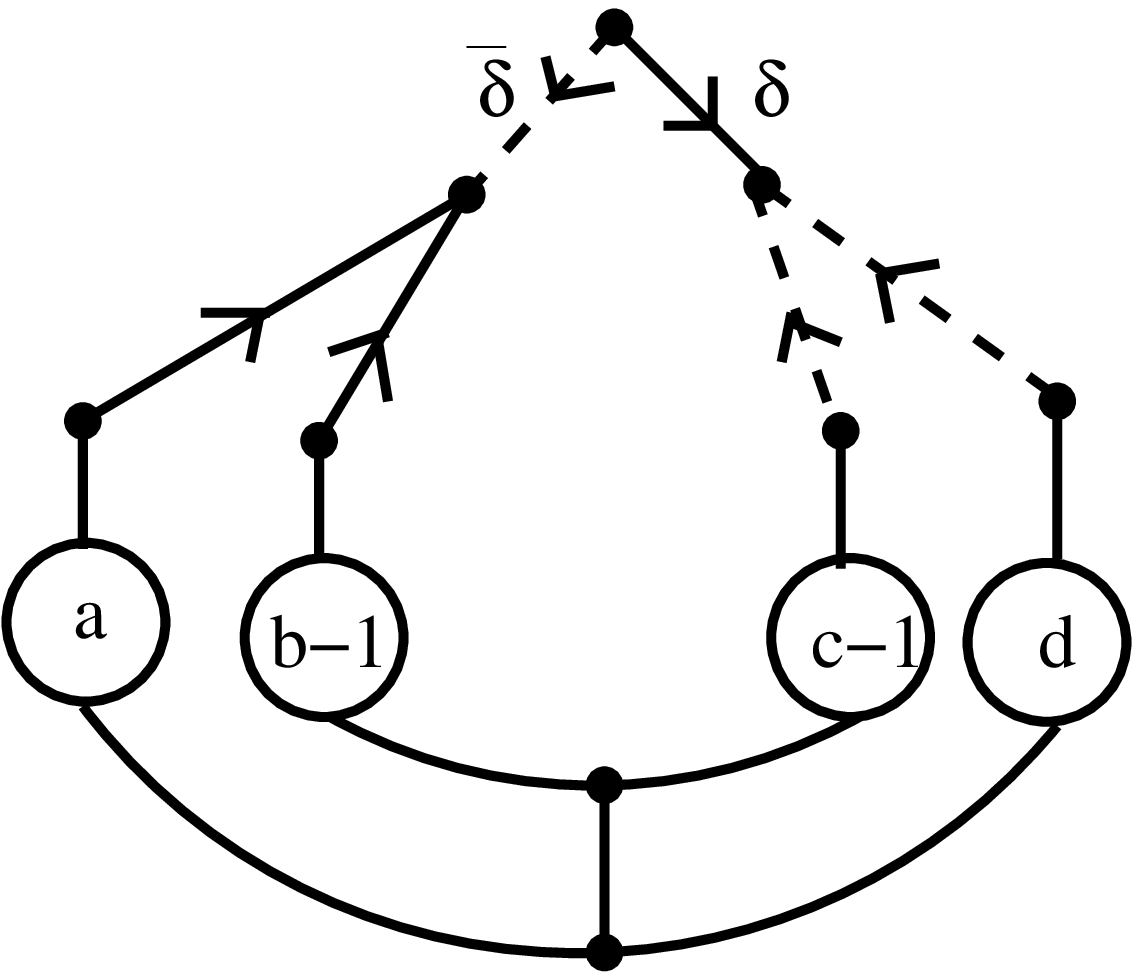}
\end{picture}}
\]
\caption{$R_2 (a;b \vert c,d)$}
\end{center}
\end{figure}

Now we get that
\bea \label{s2} S_2 (a;b) + c.c. = C_{1,a,b+1} + C_{1,a+1,b} + E_{a+b+2}- E_{a+1} E_{b+1}.\eea

\begin{figure}[ht]
\begin{center}
\[
\mbox{\begin{picture}(180,110)(0,0)
\includegraphics[scale=.6]{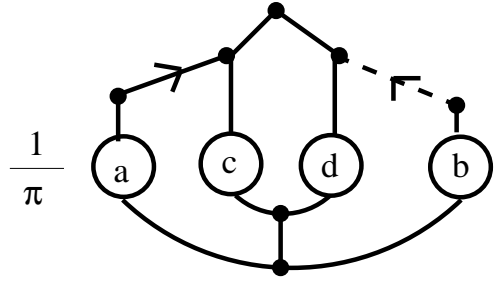}
\end{picture}}
\]
\caption{$S_3 (a,b ; c,d)$}
\end{center}
\end{figure}

\begin{figure}[ht]
\begin{center}
\[
\mbox{\begin{picture}(180,120)(0,0)
\includegraphics[scale=.6]{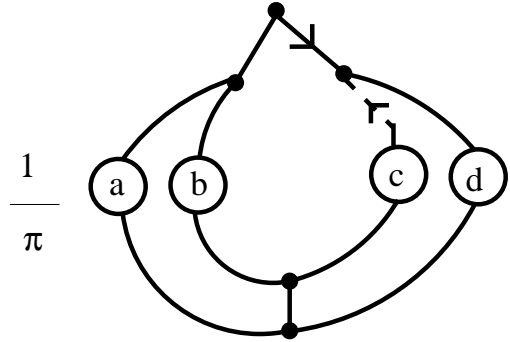}
\end{picture}}
\]
\caption{$S_4 (a,b ; c,d)$}
\end{center}
\end{figure}

We next consider the graph $S_1 (a,b \vert c,d)$. Starting with the auxiliary graph $R_2 (a,b\vert c,d)$ given in figure 30, we get that
\bea \label{S1}S_1 (a, b\vert c,d) &=& S_3 (a,d; b-1,c-1) + S_3 (b-1,c-1;a,d) \non \\ &&-\frac{1}{2}\Big[ S_3 (a-1,d;b,c-1) + S_3 (b-2,c-1;a+1,d) + S_3 (a,d-1;b-1,c) \non \\ &&+ S_3 (b-1,c-2;a,d+1)\Big]+\frac{1}{2} \Big[ S_4 (a+1,b-1;c-1,d) \non \\ &&+ S_4 (b,a ; d,c-1) + S_4 (c,d;a,b-1)^* + S_4 (d+1,c-1;b-1,a)^*\Big],\eea
where $S_3(a,b;c,d) = S_3 (b,a;d,c)^*$ and $S_4(a,b;c,d)$ are defined in figures 31 and 32 respectively.

\begin{figure}[ht]
\begin{center}
\[
\mbox{\begin{picture}(180,120)(0,0)
\includegraphics[scale=.75]{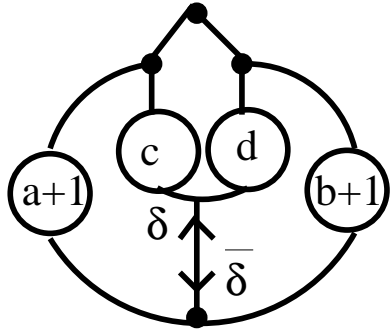}
\end{picture}}
\]
\caption{$R_3 (a,b ; c,d)$}
\end{center}
\end{figure}

For $S_3 (a,b;c,d) + c.c.$, we consider the auxiliary graph $R_3 (a,b;c,d)$ given in figure 33, leading to
\bea \label{s3}S_3 (a,b;c,d) + c.c. = M_{2,a,b+1,c,d,1} + M_{2,a+1,b,c,d,1} + C_{2,a+b+2,c+d} - P_{a+1,c;b+1,d;2}.\eea

We also obtain that
\bea \label{s4}S_4 (a,b;c,d) + c.c. = M_{2,c,b,d,a,1} - M_{2,c+1,b,d-1,a,1} + M_{1,c+1,b,d,a,1}.\eea

Thus from \C{s3} and \C{s4} we see that $S_1 (a,b \vert c,d) + c.c.$ has graphs with no derivatives. Along with \C{q1}, \C{Q3}, \C{Q5}, \C{s2} we see that $K(a,b,c,d) + c.c.$ has graphs with no derivatives, once again leading to a simplification in the structure of the source terms.    

\section{The eigenvalue equation for $C_{a,b,c,d}$}

Let us now consider \C{eqn2} in detail, which is the required eigenvalue equation satisfied by $Y_{a,b,c,d}$ for generic $a,b,c$ and $d$. The right hand side of the equation involves $f(a,b;c,d)$ which can be expressed in terms of various modular graphs without derivatives using the various expressions deduced in the previous section. This involves terms linear and quadratic in $E_a$, as well as terms linear in $C_{a,b,c}$, $C_{a,b,c,d}$, $P_{a,b;c,d;e}$ and $M_{a,b,c,d,e,f}$.   

It is useful to isolate explicitly the expressions that are quadratic in $E_a$ and linear in $C_{a,b,c,d}$ from the right hand side of \C{eqn2}. This is because the eigenvalue equation involves the Laplacian acting on $Y_{a,b,c,d}$ which is a linear combination of $C_{a,b,c,d}$ and terms quadratic in $E_a$. 
  
Keeping only these terms, we see that
\bea D(a,b;c,d) + c.c. &\rightarrow &2\Big(E_{a+b+1}E_{c+d} + E_{a+b+2}E_{c+d-1}\Big) - \Big(C_{a+1,b+1,c-1,d} + C_{a+1,b+1,c,d-1}\Big) \non \\ &&+ E_{a+c}E_{b+d+1} + E_{a+d}E_{b+c+1} + E_{b+c}E_{a+d+1}+ E_{b+d} E_{a+c+1},\eea
while
\bea K(a,b;c,d) + c.c. \rightarrow C_{a+1,b+1,c,d} - E_{a+b+2}E_{c+d} - E_{a+c+1} E_{b+d+1}.\eea

Thus the contribution to $f(a,b;c,d)$ coming from these terms is given by
\bea f(a,b;c,d) &\rightarrow& ab\Big[2 C_{a,b,c,d}  + 4  C_{a+1,b+1,c-1,d-1} + \Big(C_{a+1,b+1,c-2,d} + C_{a+1,b+1,c,d-2}\Big)\non \\ &&+ 2 E_{a+b}E_{c+d} -6 E_{a+b+2} E_{c+d-2} - 4\Big(E_{a+c}E_{b+d} + E_{a+d}E_{b+c}\Big) \non \\ &&-\Big(E_{a+c-1}E_{b+d+1}+ E_{a+c+1}E_{b+d-1} + E_{a+d-1}E_{b+c+1} + E_{a+d+1}E_{b+c-1}\Big)\Big].\non \\ \eea

Thus we get the Poisson equation
\bea \label{ev}\Big[\Delta - \lambda_{a,b,c,d}\Big]Y_{a,b,c,d} = \mathcal{C}_{a,b,c,d} + \mathcal{E}^{(1)}_{a,b,c,d}-\mathcal{E}^{(2)}_{a,b,c,d} +\mathcal{F}_{a,b,c,d}.\eea
In \C{ev} the eigenvalue $\lambda_{a,b,c,d}$ is given by
\be \lambda_{a,b,c,d} = a(a-1) + b(b-1) + c(c-1) + d(d-1).\ee

Let us now consider the various source terms that arise in this Poisson equation. 
We see that $\mathcal{C}_{a,b,c,d}$ includes all terms linear in graphs of the type $C_{a,b,c,d}$ with the structure of labels distinct from that on the left hand side. This is given by
\bea \mathcal{C}_{a,b,c,d} &=& ab \Big[ 4 C_{a+1,b+1,c-1,d-1} + C_{a+1,b+1,c-2,d} + C_{a+1,b+1,c,d-2}\Big] \non \\&&+ac \Big[ 4 C_{a+1,c+1,b-1,d-1} + C_{a+1,c+1,b-2,d} + C_{a+1,c+1,b,d-2}\Big] \non \\&& +ad \Big[ 4 C_{a+1,d+1,b-1,c-1} + C_{a+1,d+1,b-2,c} + C_{a+1,d+1,b,c-2}\Big] \non \\ &&+bc \Big[ 4 C_{b+1,c+1,a-1,d-1} + C_{b+1,c+1,a-2,d} + C_{b+1,c+1,a,d-2}\Big] \non \\&& +bd \Big[ 4 C_{b+1,d+1,a-1,c-1} + C_{b+1,d+1,a-2,c} + C_{b+1,d+1,a,c-2}\Big] \non \\ &&+cd \Big[ 4 C_{c+1,d+1,a-1,b-1} + C_{c+1,d+1,a-2,b} + C_{c+1,d+1,a,b-2}\Big]  .\eea
Hence this contribution includes terms from the same family as the parent graph $C_{a,b,c,d}$ but with different labels. Qualitatively they are on the same footing as the source terms in the Poisson equations that have been obtained for $C_{a,b,c}$ and $M_{a,b,c,d,ef}$~\cite{DHoker:2015gmr,Kleinschmidt:2017ege}. The remaining source terms are qualitatively different, which we now discuss. 
 
In \C{ev} $\mathcal{E}^{(1)}_{a,b,c,d}$ includes all terms linear in graphs of the type $E_{a}$ and is given by
\bea\mathcal{E}^{(1)}_{a,b,c,d} = 8\Big(ab+ac+ad+bc+bd+cd\Big)E_{a+b+c+d} ,\eea
while $\mathcal{E}^{(2)}_{a,b,c,d}$ includes all terms quadratic in graphs of the type $E_a$ and is given by
\bea \mathcal{E}^{(2)}_{a,b,c,d} &=& (a+b)(c+d) \Big[4E_{a+b}E_{c+d}+ E_{a+b+1}E_{c+d-1} + E_{a+b-1}E_{c+d+1}\Big] \non \\&&+ (a+c)(b+d)\Big[ 4 E_{a+c}E_{b+d}+ E_{a+c+1}E_{b+d-1} + E_{a+c-1}E_{b+d+1}\Big]\non \\ &&+(a+d)(b+c)\Big[ 4E_{a+d}E_{b+c} + E_{a+d+1}E_{b+c-1}+ E_{a+d-1}E_{b+c+1}\Big] \non\\ &&+ 6\Big[ab E_{a+b+2}E_{c+d-2}+ acE_{a+c+2}E_{b+d-2}+ ad E_{a+d+2}E_{b+c-2}\non \\ &&+ bc E_{b+c+2}E_{a+d-2}+ bd E_{b+d+2}E_{a+c-2}+ cd E_{c+d+2}E_{a+b-2}\Big].\eea

Finally, $\mathcal{F}_{a,b,c,d}$ includes the total contribution coming from the graphs of the type $C_{a,b,c}$, $P_{a,b;c,d;e}$ and $M_{a,b,c,d,e,f}$. It is given by
\bea \mathcal{F}_{a,b,c,d} &=& F(a,b;c,d) + F(a,c;b,d) + F(a,d;b,c) + F(b,c;a,d)\non \\ && + F(b,d;a,c) + F(c,d;a,b),  \eea
where
\bea \label{longexp}F(a,b;c,d) &=& ab\Big[  \mathcal{K}(c-1,b;a,d-1)+ \mathcal{K}(d-1,b;a,c-1)+\mathcal{K}(c-1,d-1;a,b) \non \\ &&+ \mathcal{K}(d-1,c-1;a,b)+\mathcal{K}(a,b;c-1,d-1)+ \mathcal{K}(a,b;d-1,c-1) \non \\ && + \mathcal{K}(a,c-1;d-1,b)+ \mathcal{K}(a,d-1;c-1,b)  + \mathcal{D}(a-1,b;c,d) \non \\ &&+ \mathcal{D}(a,b-1;c,d) - \mathcal{D}(a,b;c-1,d) - \mathcal{D}(a,b;c,d-1)\Big] \non \\ &=& F(b,a;c,d) = F(a,b;d,c) = F(b,a;d,c),\eea
where
\bea  \mathcal{D}(a,b;c,d) = D(a,b;c,d) + c.c., \quad  \mathcal{K}(a,b;c,d) = K(a,b;c,d)+ c.c.\eea 
on comparing with \C{deff} and keeping only the relevant terms.
Plugging in the various expressions, we have that
\bea \mathcal{D}(a,b;c,d)&=&2\Big(C_{1,a+c,b+d} + C_{1,a+d,b+c} -C_{1,a+b,c+d} - C_{1,a+b+2,c+d-2} \Big) + 4 C_{1,a+b+1,c+d-1} \non \\ &&- C_{1,a+c+1,b+d-1} - C_{1,a+c-1,b+d+1} - C_{1,a+d+1.b+c-1} - C_{1,a+d-1,b+c+1} \non \\ &&+ P_{a+1,c-1;b+1,d-1;1} + P_{a+1,d-1;b+1,c-1;1} + P_{a,c;b,d;1} + P_{a,d;b,c;1} \non \\ &&-P_{a+1,c-1;b,d;1}- P_{a+1,d-1;b,c;1} - P_{a,c;b+1,d-1;1}- P_{a,d;b+1,c-1;1} \non \\ &&-M_{1,a,b+1,c-1,d-1,1} - M_{1,a+1,b,c-1,d-1,1} -M_{1,a,b+1,d-1,c-1,1} - M_{1,a+1,b,d-1,c-1,1}\non \\ &&-M_{1,c-1,d,a,b,1}- M_{1,c,d-1,a,b,1} -M_{1,d-1,c,a,b,1}- M_{1,d,c-1,a,b,1}\non \\ &&+M_{1,a+1,b-1,c-1,d,1}+ M_{1,a-1,b+1,c,d-1,1} +M_{1,a+1,b-1,d-1,c,1}+ M_{1,a-1,b+1,d,c-1,1}\non \\&&+ M_{1,c-2,d,a+1,b,1} + M_{1,c,d-2,a,b+1,1}+ M_{1,d-2,c,a+1,b,1} + M_{1,d,c-2,a,b+1,1}, \non \\ &=&\mathcal{D}(b,a;c,d) =\mathcal{D}(a,b;d,c) = \mathcal{D}(b,a;d,c), \eea
and
\bea \mathcal{K} (a,b;c,d)&=& 2 \Big(C_{1,a+b+1,c+d} +  C_{1,a+b+2,c+d-1} +  C_{2,a+b+1,c+d-1}\Big)\non \\ &&-C_{2,a+b+2,c+d-2}- C_{2,a+b,c+d} -P_{a+1,c-1;b+1,d;1} - P_{a+1,c;b+1,d-1;1} \non \\ &&-P_{a+1,c-1;b,d;2} + P_{a+1,c-1;b+1,d-1;2} + P_{a,c;b,d;2} - P_{a,c;b+1,d-1;2} \non \\ &&- M_{2,a,b+1,c-1,d-1,1} - M_{2,a+1,b,c-1,d-1,1}+M_{2,a,b+1,c,d-2,1}+ M_{2,a+1,b,c-2,d,1}\non \\ &&+M_{2,a-1,b+1,c,d-1,1}+ M_{2,a+1,b-1,c-1,d,1}- M_{2,a,b,c,d-1,1} - M_{2,a,b,c-1,d,1}\non \\ &&+ M_{1,a+1,b,c,d-1,1}- M_{1,a+1,b,c-1,d,1} + M_{1,a,b+1,c-1,d,1}- M_{1,a,b+1,c,d-1,1} \non \\ &=& \mathcal{K} (b,a;d,c).\eea

Thus \C{ev} gives the Poisson equation satisfied by $C_{a,b,c,d}$ for generic values of $a,b,c$ and $d$. Now formally $C_{a,b,c,d}$ is defined for all $a,b,c$ and $d$ so long as the integrand has nice convergence properties, though such graphs arise in string amplitudes for $a,b,c$ and $d$ being positive integers\footnote{In the expression for the string amplitude when the labels are positive integers, the integrand involving the modular graph has to be integrated over the truncated fundamental domain of $SL(2,\mathbb{Z})$. For this integral to make sense, the graph must have appropriate convergence properties when expanded around the cusp. This needs a detailed analysis of the asymptotic expansion. This should generalize to generic choices of the labels. For $C_{a,b,c,d}$ it will be interesting to perform this analysis along the lines of~\cite{DHoker:2019txf}. Similar issues have been discussed at genus two, for example, in~\cite{DHoker:2013fcx}.}. Thus from the structure of the source terms, we see that there can be contributions involving expressions with vanishing or negative labels. We now analyze some aspects of such graphs.

First let us consider the case where $a,b,c,d >2$. There are no graphs with negative labels, while the only graphs with vanishing labels are of the type $M_{1,0,a,b,c,1}$ and $M_{2,0,a,b,c,1}$ for $a,b,c >0$ which arise when one of the labels in the parent graph is 3. While $M_{1,0,a,b,c,1}$ arises from terms of the form $\mathcal{D}(a,b,c,d)$, $M_{2,0,a,b,c,1}$ arises from terms of the form $\mathcal{K}(a,b,c,d)$. Now $M_{1,0,a,b,c,1}$ and $M_{2,0,a,b,c,1}$ are given as special cases of the first equation in \C{M1}. Thus along with these definitions, we see that \C{ev} is the eigenvalue equation for all $a,b,c,d >2$. Graphs with vanishing or negative labels have an interpretation as degeneration limits of graphs with certain insertions along certain links, as explained in the appendix. Thus it is expected that the structure of the eigenvalue equation can qualitatively change when the source terms have such graphs. From \C{M1} we see how this happens, as we are left with source terms of the form $P_{a,b;c,d;e}$ and $C_{a,b,c}$ with positive labels rather than terms of the form $M_{1,0,a,b,c,1}$ and $M_{2,0,a,b,c,1}$. However, the change is not drastic as the eigenvalue does not change.       

The remaining graphs $C_{a,b,c,d}$ are given by: (i) $C_{a,b,c,d}$ where $a,b,c,d >1$ and at least one label is 2, and (ii) $C_{a,b,c,d}$ where at least one label is 1. 
The eigenvalue equation for each of them yields source terms that have graphs with vanishing as well as negative labels.

We first consider the graphs $C_{a,b,c,d}$ of type (i) where $a,b,c,d>1$ and at least one label is 2. From \C{ev}, we see that there are several contributions which involve vanishing labels of the type given by 
\be C_{2,a,0}, C_{a,b,c,0}, P_{a,0;b,c;1}, P_{a,0;b,c;2}, P_{a,0;b,0;2}, M_{1,0,a,b,c,1}, M_{2,0,a,b,c,1}, M_{2,a,b,0,0,1}, M_{2,0,a,b,0,1}\ee
where $a,b,c>0$, which are evaluated in the appendix. Note that $M_{2,0,a,b,0,1} = C_{a,b,1,2}+\ldots,$ and hence can change the eigenvalue in \C{ev} leading to a drastic change in  the structure of the equation. Apart from this, there are also contributions which involve negative labels  of the type given by
\be \label{zerosim} M_{1,-1,a,b,c,1}, M_{2,-1,a,b,c,1}\ee
where $a,b,c>0$. While such contributions are cumbersome to simplify and evaluate, let us extract the origin of such problematic terms.   

We obtained the eigenvalue equation for generic values of the labels of the original graph, and vanishing or negative labels arise for particular values of the labels of certain graphs that arise as the source terms. This is simply because  at a certain intermediate step of the analysis, the label in a the link of a graph $\mathcal{G}$ vanishes. Continuing with this graph further, yields vanishing or negative labels. Thus to obtain answers with positive labels only, one has to stop at that stage of the analysis and simplify the graph $\mathcal{G}$, obtain graphs with only positive labels~\footnote{How to do so for any graph is explained in the appendix.}, and then proceed. Doing this at every relevant step of the analysis yields graphs in the eigenvalue equation involving only positive labels.

It is easy to see how this works for \C{zerosim}. Retracing back several steps, we see that such a contribution comes from $\mathcal{D}(a,b;c,d) = -Q_2 (a,c\vert d,b) - Q_2 (a,d\vert c,b) + c.c. +\ldots$, on using
\bea Q_2 (a,b\vert c,d) &=& -\frac{1}{2}\Big[  Q_3 (b-2,c-1;a+1,d) + Q_3 (b-1,c-2;a,d+1)\Big] \non \\ 
 &&+\frac{1}{2} \Big[  Q_5 (b,a ; d,c-1) + Q_5 (c,d;a,b-1)^* \Big]+\ldots.\eea 
This arises from \C{Q2} where we have kept only the relevant terms. Similarly such a contribution comes from $\mathcal{K}(a,b;c,d) = -S_1 (a,c\vert d,b) + c.c. +\ldots$, on using
\bea S_1 (a,b\vert c,d) &=& -\frac{1}{2}\Big[  S_3 (b-2,c-1;a+1,d) + S_3 (b-1,c-2;a,d+1)\Big] \non \\ &&+\frac{1}{2} \Big[  S_4 (b,a ; d,c-1) + S_4 (c,d;a,b-1)^* \Big]+\ldots\eea 
which arises from \C{S1}. Thus in $F(a,b,2,d)$, such terms arise from $Q_2 (a,1\vert,  d,b)$, $Q_2 (a,d\vert 1,b)$ and $S_1 (a,1\vert d,b)$ where $a,b,d >1$. From figures 20 and 28, this is precisely the case when one of the labels in one of the links vanishes. Thus for these cases, one should proceed with the graphs given in figure 34 instead, leading to source terms with only positive labels.  

\begin{figure}[ht]
\begin{center}
\[
\mbox{\begin{picture}(350,150)(0,0)
\includegraphics[scale=.65]{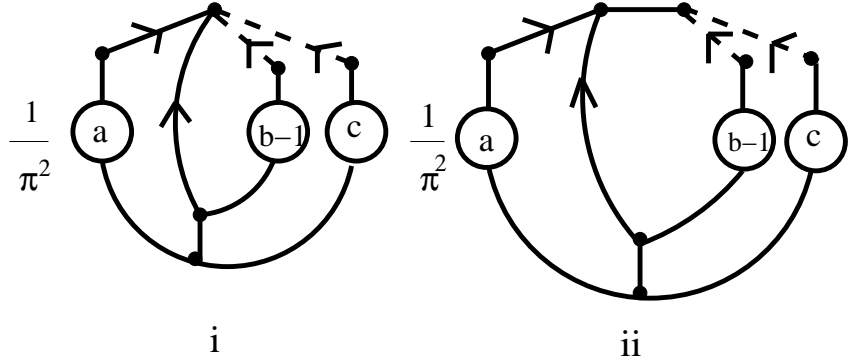}
\end{picture}}
\]
\caption{(i) $Q_2 (a,1\vert b,c)$ and (ii) $S_1 (a,1\vert b,c)$}
\end{center}
\end{figure}

Now when one considers graphs of the form (ii), there are more such contributions to consider. In fact, the source terms also yield $E_1$\footnote{This is logarithmically divergent if evaluated with a momentum cutoff and has to be regularized.} which must be absent in the final expression. This is because from \C{Ea} we see this arises when one Green function forms a closed loop. This divergent contribution is not part of the moduli space which defines modular graphs, and must be excluded.  
 
Thus we see that for cases (i) and (ii) there are several cumbersome contributions which arise whose total contributions must yield graphs with positive labels only in the eigenvalue equation. It will be interesting to analyze these cases for these choices of labels.

Since there many such problematic contributions, it is useful to instead take a direct approach to calculating the eigenvalue equation which seems easier to implement. To do so, we start with \C{eqn2} instead and proceed directly. In the appendix, we obtain the eigenvalue equations for $C_{1,1,1,1}$, $C_{1,1,1,2}$ and $C_{1,1,2,2}$ this way. This circumvents the issues discussed above, and yields the final answer directly.   

Thus we have obtained the Poisson equation satisfied by the family of modular graphs $C_{a,b,c,d}$ for generic values of $a,b,c$ and $d$. We have also analyzed in some detail the issue of graphs with vanishing or negative labels that arise as source terms for certain specific labels of the parent graph. This analysis based on introducing various auxiliary graphs should generalize to other families of modular graphs leading to eigenvalue equations. It will also be interesting to explore similar features of genus two modular graphs, where very few eigenvalue equations are understood~\cite{DHoker:2014oxd,Basu:2018bde}.      

\vspace{.5cm}

{\bf{Acknowledgements:}} I am thankful to Santosh Rai for useful comments on the draft. 

\section{Appendix}

\appendix

\section{The modular graphs with vanishing or negative labels}

In the main text, we saw that for the physically interesting case of string amplitudes, one can have source terms in the Poisson equations involving modular graphs with vanishing or negative labels. We now analyze the content of several modular graphs of such kind. We shall see that such graphs make perfect sense and such labels imply degeneration limits of the graphs, which can be analyzed directly. 

To analyze such graphs, let us consider writing them as lattice sums resulting in purely algebraic expressions that result from using \C{exp} and integrating over the vertices. First we consider the case where  a link has a vanishing label. This simply amounts to inserting a factor of $\pi^{-1} \p_z \overline\p_w G(z,w)$ along that link. From \C{exp} we see this produces just 1 as the factor corresponding to that link which precisely yields the link with vanishing label. Generalizing this, it follows that a link with label $-n$ where $n$ is a non--negative integer corresponds to inserting a factor of $\pi^{-(n+1)} \p_z^{n+1} \overline\p_w^{n+1} G(z,w)$ along that link. 

 Thus we see that graphs with vanishing or negative labels come with these factors inserted along those links. Hence for $n=0$ they can be simplified using \C{Eigen}, while for $n >0$ they can be simplified using \C{Eigen} as well as integrating by parts. Thus it follows that graphs with vanishing or negative labels are degenerate limits of the parent graphs associated with the joining and splitting of vertices. Consequently they are topology changing processes. We now analyze several cases that are relevant for our purposes.   

\begin{figure}[ht]
\begin{center}
\[
\mbox{\begin{picture}(300,140)(0,0)
\includegraphics[scale=.75]{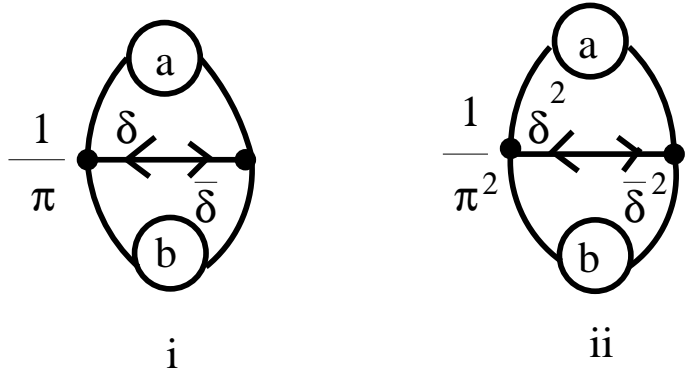}
\end{picture}}
\]
\caption{(i) $C_{a,b,0}$ and (ii) $C_{a,b,-1}$}
\end{center}
\end{figure}

\subsection{The graphs $C_{a,b,0}$ and $C_{a,b,-1}$}

We first consider $C_{a,b,0}$ and $C_{a,b,-1}$. Based on the above arguments, 
they are given in figure 35. Thus $C_{a,b,0}$ can be evaluated by using \C{Eigen} directly, while evaluating $C_{a,b,-1}$ also involves integrating a factor of $\p$ and a factor of $\overline\p$ by parts. We get that
\bea C_{a,b,0} &=& E_a E_b - E_{a+b}, \non \\ 
 C_{a,b,-1} &=& E_a E_{b-1} + E_{a-1}E_b.\eea
They have been evaluated in~\cite{DHoker:2015gmr} using the lattice representation of the graphs. 

\begin{figure}[ht]
\begin{center}
\[
\mbox{\begin{picture}(300,150)(0,0)
\includegraphics[scale=.65]{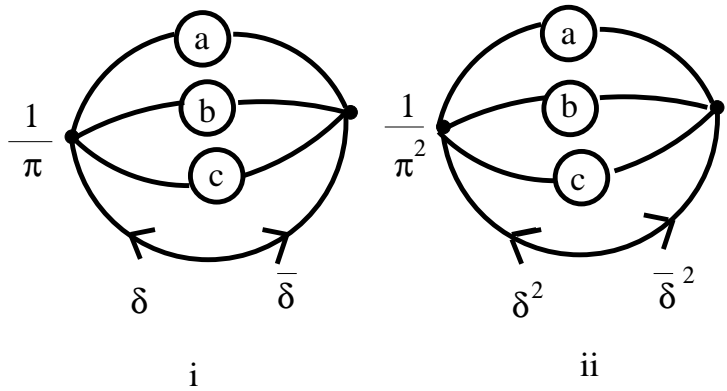}
\end{picture}}
\]
\caption{(i) $C_{a,b,c,0}$ and (ii) $C_{a,b,c,-1}$}
\end{center}
\end{figure}

\subsection{The graphs $C_{a,b,c,0}$ and $C_{a,b,c,-1}$}

We next consider $C_{a,b,c,0}$ and $C_{a,b,c,-1}$ which are given in figure 36, and evaluate to

\bea  C_{a,b,c,0} &=& E_a E_b E_c - C_{a,b,c}, \non \\
 C_{a,b,c,-1} &=& E_{a-1} E_b E_c + E_a E_{b-1} E_c + E_a E_b E_{c-1}.\eea

\begin{figure}[ht]
\begin{center}
\[
\mbox{\begin{picture}(340,340)(0,0)
\includegraphics[scale=.75]{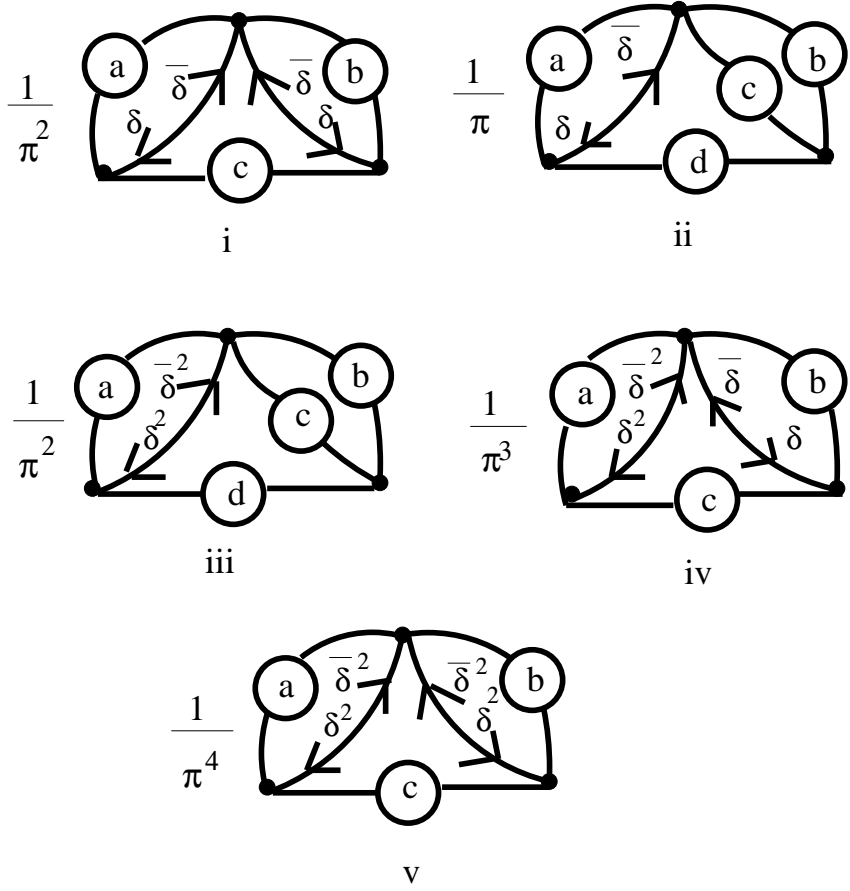}
\end{picture}}
\]
\caption{(i) $P_{a,0;b,0;c}$, (ii) $P_{a,0;b,c;d}$, (iii) $P_{a,-1;b,c;d}$, (iv) $P_{a,-1;b,0;c}$ and (v) $P_{a,-1;b,-1;c}$}
\end{center}
\end{figure}

\subsection{The graphs $P_{a,0;b,0;c}$, $P_{a,0;b,c;d}$, $P_{a,-1;b,c;d}$, $P_{a,-1;b,0;c}$ and $P_{a,-1;b,-1;c}$}

The graphs $P_{a,0;b,0;c}$, $P_{a,0;b,c;d}$, $P_{a,-1;b,c;d}$, $P_{a,-1;b,0;c}$ and $P_{a,-1;b,-1;c}$ are given in figure 37. Proceeding as above, they yield

\bea P_{a,0;b,0;c} &=& E_a C_{b,c,0} - C_{a+c,b,0} = E_{a+b+c}- E_a E_{b+c} - E_{a+c} E_b + E_a E_b E_c, \non \\
 P_{a,0;b,c;d} &=& E_a C_{b,c,d} - C_{a+d,b,c}, \non \\
P_{a,-1;b,c;d} &=& E_{a-1} C_{b,c,d} + E_a C_{b,c,d-1} , \non \\
P_{a,-1;b,0;c} &=& E_b C_{a,c,-1} - C_{a,b+c,-1} = E_{a-1} E_{b} E_c + E_a E_b E_{c-1}- E_{a-1}  E_{b+c} - E_a E_{b+c-1}, \non \\
P_{a,-1;b,-1;c} &=& E_{a-1} C_{b,c,-1} + E_a C_{b,c-1,-1} = E_a E_b E_{c-2} + E_{a-1}E_{b-1}E_c\non \\ &&+ (E_{a-1}E_b + E_a E_{b-1})E_{c-1}. \eea

\begin{figure}[ht]
\begin{center}
\[
\mbox{\begin{picture}(340,280)(0,0)
\includegraphics[scale=.7]{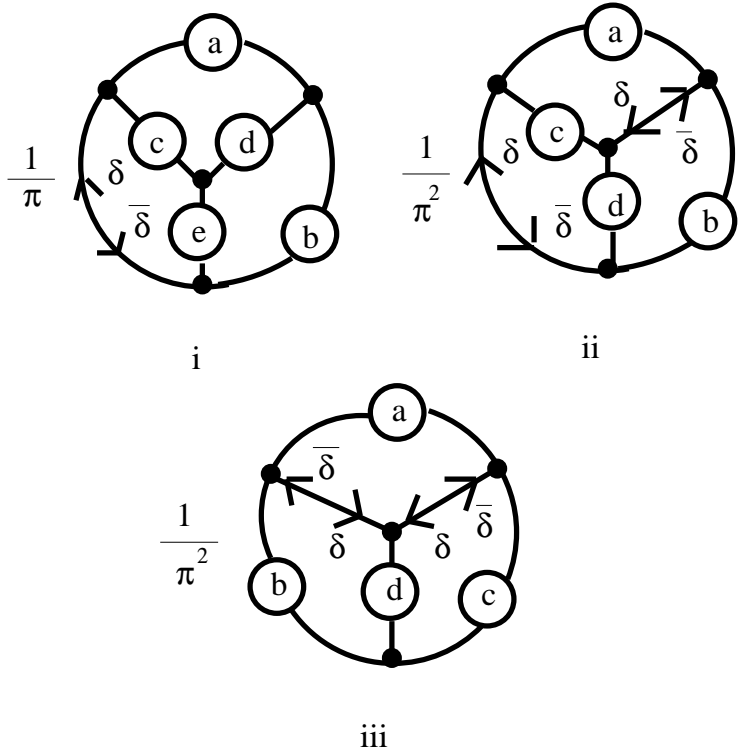}
\end{picture}}
\]
\caption{(i) $M_{a,0,b,c,d,e}$, (ii) $M_{a,0,b,c,0,d}$ and (iii) $M_{a,b,c,0,0,d}$}
\end{center}
\end{figure}

\subsection{The graphs $M_{a,0,b,c,d,e}$, $M_{a,0,b,c,0,d}$ and $M_{a,b,c,0,0,d}$}
Finally we consider the graphs $M_{a,0,b,c,d,e}$, $M_{a,0,b,c,0,d}, M_{a,b,c,0,0,d}$, $M_{a,-1,b,c,d,e}$ which are given in figure 38. They evaluate to 
\bea \label{M1}M_{a,0,b,c,d,e} &=& P_{a,b;c,e;d} - C_{a+c,b+e,d}, \non \\
M_{a,0,b,c,0,d}&=& C_{a,b,c,d} + E_{a+b+c+d} -E_{a+b}E_{c+d} - E_{a+c}E_{b+d}, \non \\
 M_{a,b,c,0,0,d} &=& E_a C_{b,c,d} - C_{a+c,b,d} -C_{a+b,c,d}.\eea
Some of these relations have been derived in~\cite{Kleinschmidt:2017ege} using the lattice representation of the graph. 

Thus for all these cases we see that we are left with various modular graph functions where the various links are given by scalar Green functions.   

One can analyze other contributions along these lines. Since the basic argument for this graphical technique is very general, this must hold true for arbitrary graphs with vanishing or negative labels and hence can be used to simplify them.    

\section{The eigenvalue equations for $C_{1,1,1,1}$, $C_{1,1,1,2}$ and $C_{1,1,2,2}$}

The eigenvalue equation for $C_{1,1,1,1}$, $C_{1,1,1,2}$ and $C_{1,1,2,2}$ in \C{ev} yields several source terms with vanishing and negative labels. Hence we start directly with \C{eqn2} instead and obtain the eigenvalue equations\footnote{Note that for $a,b,c,d \geq 1$, $f(a,b;c,d)$ has no terms with negative labels.}. We calculate terms of the form $D(a,b;c,d)$ and $K(a,b;c,d)$ using the techniques in the main text, and give only the final answers for the sake of brevity.  

\subsection{The eigenvalue equation for $C_{1,1,1,1}$}

From \C{eqn2}, we get that
\be \Delta (C_{1,1,1,1} - 3 E_2^2) +12 E_2^2 + 12 C_{1,1,1,1} = 6 f(1,1;1,1).\ee

We now calculate the various terms that arise in $f(1,1;1,1)$.  

For terms of the type $D(a,b;c,d)$, we get that
\bea &&D(0,1;1,1) = D(1,0;1,1) = \frac{C_{1,1,1,1}}{3}, \non \\ &&D(1,1;0,1) = D(1,1;1,0)= -C_{1,1,2} + \frac{1}{2} (E_2^2 - E_4),\eea
while for terms of the type $K(a,b;c,d)$ we get that
\bea && K (0,1;1,0) = K(1,0;0,1) = -\frac{C_{1,1,2}}{2}, \non \\ && K(0,0;1,1) = \frac{1}{4} (C_{1,1,1,1} - E_2^2), \quad K(1,1;0,0) = E_4. \eea
Adding the various contributions we get the eigenvalue equation~\cite{DHoker:2015gmr,DHoker:2016mwo,Basu:2016kli}
\be (\Delta -2) (C_{1,1,1,1}-3 E_2^2) = 36 E_4 - 24 E_2^2.\ee

\subsection{The eigenvalue equation for $C_{1,1,1,2}$}

Again from \C{eqn2}, we get that

\pagebreak

\be \Delta (C_{1,1,1,2} - 3 E_2 E_3) + 24 E_2 E_3 + 16 C_{1,1,1,2} = 3\Big[f(1,1;1,2) + f(1,2;1,1)\Big].\ee

While terms of the type $D(a,b;c,d)$ give us
\bea &&D(0,2;1,1) = \frac{C_{1,1,1,2}}{3}, \quad D (1,1;0,2) = -\frac{C_{1,2,2}}{2}, \non \\ && D(1,2;0,1) + c.c.=  -C_{1,1,3} -C_{1,2,2} + E_2 E_3 - E_5, \non \\ && 2 D(1,0;1,2) + D(1,1;1,1) =  C_{1,1,1,2},\eea
terms of the type $K(a,b;c,d)$ give us
\bea &&K(0,2;1,0) = -\frac{C_{1,1,3}}{2}, \quad K (1,2;0,0) = E_5, \non \\&&
K(0,1;1,1) + c.c. = \frac{1}{2} (C_{1,1,1,2} - E_2 E_3 + P_{1,1;1,1;1} + C_{1,1,3} - E_2 C_{1,1,1}),\non \\&& K(1,1;1,0) + c.c. = E_2 E_3 - E_5 - C_{1,2,2} - C_{1,1,3}, \non \\ &&K(0,0;1,2) + c.c. = \frac{1}{2} (C_{1,1,1,2} - E_2 E_3- P_{1,1;1,1;1} - C_{1,1,3} + E_2 C_{1,1,1}),\non \\ &&K(1,0;0,2) + c.c. = C_{1,1,3}  -C_{1,2,2} -E_2 E_3 + E_5.\eea
Adding the various contributions we get the eigenvalue equation~\cite{DHoker:2015gmr,DHoker:2016mwo,Basu:2016kli}
\bea  (\Delta -6) (C_{1,1,1,2} - 3 E_2 E_3) = 36 E_5 - 9 C_{1,2,2} -30 E_2 E_3. \eea

\subsection{The eigenvalue equation for $C_{1,1,2,2}$}

Once again from \C{eqn2}, we get that
\bea &&\Delta (C_{1,1,2,2} -  E_2 E_4 - 2 E_3^2) + 14 E_2 E_4 + 24 E_3^2 +22 C_{1,1,2,2} \non \\ &&= f(1,1;2,2) + f(2,2;1,1) + 4 f(1,2;1,2). \eea

Terms of the type $D(a,b;c,d)$ give us
\bea D(1,0;2,2) &=& \frac{1}{3} (E_2^3 - C_{2,2,2}), \quad D(2,2;0,1) = \frac{1}{2}(E_3^2 - E_6) - C_{1,2,3}, \non \\ D (1,2;0,2) &=& -\frac{C_{2,2,2}}{2}, \quad D(1,1;1,2) = \frac{C_{1,1,2,2}}{2} +\frac{1}{6}(C_{2,2,2} - E_2^3),\non \\ D(0,2;1,2) +c.c.&=& E_6 - 3 C_{1,2,3} + C_{1,1,2,2} - 2 C_{1,1,4} - C_{2,2,2} -P_{1,1;1,2;1}\non \\ &&+\frac{1}{2}(P_{1,1;1,1;2} - C_{1,1,1,3} + E_2 E_4 + C_{1,1,1} E_3 + E_2 C_{1,1,2}).\eea
The total contribution from $D(2,1;1,1) + c.c.$ cancels. 

Also terms of the type $K(a,b;c,d)$ give us
\bea && K(1,2;1,0) + K(0,2;2,0) = -C_{1,2,3}, \quad K(0,1;1,2) = \frac{1}{4}(C_{1,1,2,2} - E_2 E_4),\non \\ && K(1,1;1,1) + K(0,0;2,2) + 2 K(1,0;1,2) + c.c. = 2(C_{1,1,2,2} - E_3^2), \non \\ && K(2,2;0,0) = E_6, \quad K(1,1;0,2) = -\frac{C_{2,2,2}}{2},\non \\ && K(1,2;0,1) = \frac{1}{2}(E_3^2 - E_6) - C_{1,2,3},  \non \\ && K(0,2;1,1) + c.c.= \frac{1}{2} (P_{1,1;1,2;1} + C_{1,1,1,3} + C_{1,1,4}- E_2 E_4 - C_{1,1,1} E_3).\eea

Adding the various contributions we get the eigenvalue equation
\bea \label{consis}&&(\Delta -12)(C_{1,1,2,2} - E_2 E_4 - 2 E_3^2) = 24 E_6 -4 P_{1,1;1,2;1} + 4 P_{1,1;1,1;2} \non \\ &&-56 C_{1,2,3} -12C_{1,1,4} - \frac{22}{3} C_{2,2,2}  + 4 E_2 C_{1,1,2} - 16 E_3^2 - 6 E_2 E_4 - \frac{2}{3} E_2^3.\eea

Let us now make a non--trivial consistency check of \C{consis}. The various modular graphs involved in \C{consis} can be expanded around the cusp $\tau_2 \rightarrow \infty$ , and each of them yields terms that are power behaved as well as exponentially suppressed in $\tau_2$. The power behaved terms are given by~\cite{Green:2008uj,DHoker:2015gmr,Zerbini:2015rss}

\bea \label{sub}&&E_2 = \frac{y^2}{45} + \frac{\zeta(3)}{y}, \quad E_3 = \frac{2y^3}{945}+ \frac{3\zeta(5)}{4y^2}, \quad E_4 = \frac{y^4}{4725}+ \frac{5\zeta(7)}{8y^3}, \non \\&& E_6 = \frac{1382y^6}{638512875}+ \frac{63\zeta(11)}{128 y^5}, \quad  
C_{1,1,2} = \frac{2y^4}{14175} +\frac{\zeta(3)y}{45}+ \frac{5\zeta(5)}{12y} - \frac{\zeta(3)^2}{4y^2} +\frac{9\zeta(7)}{16y^3}, \non \\ &&C_{1,2,3} = \frac{43y^6}{58046625} + \frac{\zeta(5)y}{630}+ \frac{\zeta(7)}{144y} +\frac{7\zeta(9)}{64y^3} -\frac{17\zeta(5)^2}{64y^4} +\frac{99\zeta(11)}{256y^5},\non \\ &&C_{2,2,2} = \frac{38y^6}{91216125} + \frac{\zeta(7)}{24y} -\frac{7\zeta(9)}{16y^3} +\frac{15\zeta(5)^2}{16y^4} -\frac{81\zeta(11)}{128y^5}, \non \\ &&C_{1,1,4} = \frac{808y^6}{638512875} +\frac{\zeta(3)y^3}{4725}- \frac{\zeta(5)y}{1890}+ \frac{\zeta(7)}{720y} +\frac{23\zeta(9)}{64y^3} \non \\ &&-\frac{\zeta(5)^2+30\zeta(3)\zeta(7)}{64y^4} +\frac{167\zeta(11)}{256y^5}, \non \\ &&C_{1,1,2,2} = \frac{103y^6}{13030875} + \frac{\zeta(3)y^3}{2025}+ \frac{\zeta(5)y}{54} -\frac{\zeta(3)^2}{90} -\frac{\zeta(7)}{360y} +\frac{5\zeta(3)\zeta(5)}{12y^2}\non \\ &&+\frac{5\zeta(9) -48 \zeta(3)^3}{288 y^3} +\frac{14\zeta(3)\zeta(7)+ 25 \zeta(5)^2}{32 y^4} - \frac{73\zeta(11)}{128y^5}, \non \\ &&P_{1,1;1,1;2} = \frac{68y^6}{70945875} + \frac{4\zeta(3)y^3}{14175}- \frac{\zeta(5)y}{945} +\frac{\zeta(3)^2}{45} +\frac{13\zeta(7)}{45y} -\frac{5\zeta(3)\zeta(5)}{6y^2} \non  \eea

\bea&&+\frac{61\zeta(9) +12\zeta(3)^3}{36 y^3} -\frac{3\zeta(3)\zeta(7)+ \zeta(5)^2}{2y^4}+\frac{81\zeta(11)}{64y^5}, \non \\ &&P_{1,1;1,2;1} = \frac{802y^6}{638512875} + \frac{2\zeta(3)y^3}{14175}+ \frac{43\zeta(5)y}{3780} -\frac{\zeta(3)^2}{180} +\frac{11\zeta(7)}{180y} +\frac{5\zeta(3)\zeta(5)}{24y^2} \non \\ &&-\frac{65\zeta(9) +48\zeta(3)^3}{576 y^3} -\frac{6\zeta(3)\zeta(7)+ \zeta(5)^2}{64y^4} +\frac{147\zeta(11)}{256y^5}, \eea
where $y=\pi\tau_2$. Substituting the various expressions from \C{sub} in \C{consis} we find perfect agreement.


\providecommand{\href}[2]{#2}\begingroup\raggedright\endgroup

\end{document}